\documentclass{ws-ijmpe}
\usepackage[super,compress]{cite}

\usepackage{bm}
\usepackage{graphicx}

\newcommand{\beq}{\begin{equation}}
\newcommand{\eeq}{\end{equation}}
\newcommand{\beqa}{\begin{eqnarray}}
\newcommand{\eeqa}{\end{eqnarray}}

\newcommand{\be}{\begin{eqnarray}}
\newcommand{\ee}{\end{eqnarray}}
\newcommand{\nn}{\nonumber \\ }

\begin{document}

\markboth{R. Machleidt}{Historical perspective and future prospects for nuclear interactions}

\catchline{}{}{}{}{}
%%%%%%%%%%%%%%%%%%%%%%%%%%%%%%%%%%%%%%%%%%%%%%%%%%%%%%%%%%%%%%%%%%%%

\title{Historical perspective and future prospects for nuclear interactions}

\author{R. Machleidt}

\address{Department of Physics, University of Idaho\\
Moscow, Idaho 83844, USA\\
machleid@uidaho.edu}

\maketitle

\begin{history}
\received{Day Month Year}
\revised{Day Month Year}
%\accepted{Day Month Year}
%\comby{(xxxxxxxxxx)}
\end{history}

\begin{abstract}
The nuclear force is the heart of nuclear physics and, thus, the significance of this force for all of nuclear physics can hardly be overstated. Research on this crucial force has by now spanned eight decades and we are still not done. I will first review the rich history of hope and desperation, which had spin-off far beyond just nuclear physics. 
Next, I will present the current status in the field which is charcterized by the application of an effective
field theory (EFT) that is believed to represent QCD in the low energy regime typical for nuclear physics.
During the past two decades, this EFT has become the favorite vehicle to derive nuclear two- and many-body forces. 
Finally, I will take a look into the future: What developments can we expect from the next decades? 
Will the 30-year cycles of new and ``better'' ideas for efficiently describing nuclear forces go on for ever, or is there hope for closure?
\end{abstract}


\ccode{PACS numbers: 13.75.Cs, 21.30.-x, 12.39.Fe}

%\tableofcontents

\section*{Introduction}
The development of a proper theory of nuclear forces has occupied the minds of some of the brightest physicists for eight decades and has been one of the main topics of physics research in the 20th century. The original idea was that the force is created by the exchange of lighter particles (than nucleons) known as mesons, and this idea gave rise to the birth of a new sub-field of modern physics, namely, (elementary) particle physics. The modern perception of the nuclear force is that it is a residual interaction (similar to the van der Waals force between neutral atoms) of the even stronger force between quarks, which is mediated by the exchange of gluons and holds the quarks together inside a nucleon.
We will subdivide the full story into four phases, for which we also state (in parentheses) the approximate time frame:
\begin{itemize}
\item
Phase I $(1930-1960)$:
Early Attempts and Pion Theories,
\item
Phase II $(1960 - 1990)$:
Meson Models,
\item
Phase III ($1990 - 2020)$:
Chiral Effective Field Theory,
\item
Phase IV ($2020 - 2050?)$ Future:
EFT Based Models(?).
\end{itemize}
Notice the pattern of 30-year cycles.
We will now tell the tale for each cycle.

\section{Phase I (1930 -- 1960): Early Attempts and Pion Theories}
\label{sec_pions}
After the discovery of the neutron by Chadwick in 1932~\cite{Cha32}, it was clear that the atomic nucleus is built up from protons and neutrons. In such a system, electromagnetic forces cannot be the reason why the constituents of the nucleus are sticking together. Therefore, the concept of strong nuclear interactions was introduced\footnote{A detailed account of this phase is presented in the excellent book by Brown and Rechenberg~\cite{BR96}.}
with Heisenberg giving it a first shot~\cite{Hei32}.
  In 1935, a theory for this new force was started by the Japanese physicist Hideki Yukawa~\cite{Yuk35}, who suggested that the nucleons would exchange particles between each other and this mechanism would create the force. Yukawa constructed his theory in analogy to the theory of the electromagnetic interaction where the exchange of a (massless) photon is the cause of the force. However, in the case of the nuclear force, Yukawa assumed that the ``force-makers'' (which were eventually called ``mesons'') carry a mass a fraction of the nucleon mass. This would limit the effect of the force to a finite range.
Similar to other theories that were floating around in the 1930's (like the Fermi-field theory~\cite{Fer34}), Yukawa's meson theory was originally meant to represent a unified field theory for all interactions in the
atomic nucleus (weak and strong, but not electromagnetic). 
But after about 1940, it was generally agreed that strong and and weak nuclear forces should be treated separately. 

Yukawa's proposal did not receive much attention
until the discovery of the muon in cosmic ray~\cite{NA37} in 1937 after which, however, the interest in meson theory exploded.
In his first paper of 1935, Yukawa had envisioned a scalar field theory, but when the spin of the deuteron ruled that out, he considered vector fields~\cite{YS37}. Kemmer considered the whole variety of non-derivative couplings for spin-0 and spin-1 fields (scalar, pseudoscalar, vector, axial-vector, and tensor)~\cite{Kem38}. By the early 1940's, the pseudoscalar theory was gaining in popularity, since it provided
a more suitable force for light nuclei. 
In 1947, a strongly interacting meson was found in cosmic ray~\cite{LOP47} and, in 1948, in the laboratory~\cite{GL48}: the isovector pseudoscalar pion with mass around 138 MeV. It appeared that, finally, the right quantum of strong interactions had been found.

Originally, the meson theory of nuclear forces was perceived as a fundamental relativistic quantum field theory (QFT), similar to quantum electrodynamics (QED), the exemplary QFT that was so successful.
In this spirit, a lot of effort was devoted to pion field theories in the early 
1950's~\cite{TMO52,BW53,Mar52,SBH55,BH55,Mor63}. Ultimately, all of these meson QFTs failed.
 In retrospect, they would have been replaced anyhow, because meson and nucleons are not elementary particles and QCD is the correct QFT of strong interactions. However, the meson field concept failed long before QCD was invented since, even 
 when considering mesons are elementary, 
the theory was beset with problems that could not be resolved. 
Assuming the renormalizable pseudoscalar ($\gamma_5$) coupling between pions and nucleons, 
gigantic virtual pair terms emerged that were not seen experimentally in
pion-nucleon ($\pi N$) or nucleon-nucleon ($NN$) scattering. Using the pseudo-vector or derivative coupling ($\gamma_5 \gamma^\mu \partial_\mu$), these pair terms were suppressed, but this type of coupling was not renormalizable~\cite{SBH55}.
Moreover, the large coupling constant ($g^2_\pi/4\pi \approx 14$) made perturbation theory useless. Last not least, the pion-exchange potential contained unmanageable singularities at short distances.

Eventually, most of these problems will be solved by imposing chiral symmetry and introducing
the concept of an effective field theory (cf.\ Phase III, Sec.~\ref{sec_eft}), but we are not there yet.

\section{Phase II (1960 -- 1990): Meson Models}
\label{sec_mesons}

Around 1960, rich phenomenlogical knowledge about the $NN$ interaction had accumulated
due to systematic measurements of $NN$ observables~\cite{Cha57} and  advances in phase shift analysis~\cite{SYM57}.  Clear evidence for 
a repulsive core and a strong spin-orbit force emerged. This lead Sakurai~\cite{Sak60} and Breit~\cite{Bre60} to postulate the existence of a neutral vector meson ($\omega$ meson), which would create both these features.
 Moreover, Nambu~\cite{Nam57} and Frazer and Fulco~\cite{FF59} showed that
a $\omega$ meson and  a 2$\pi$ $P$-wave resonance ($\rho$ meson)
would explain the electromagnetic structure of the nucleons.
Soon after these predictions, heavier (non-strange) mesons were found in experiment, notably the vector (spin-1) mesons
$\rho(770)$ and $\omega(782)$~\cite{Mag61,Erw61}.
It became now fashionable, to add these newly discovered mesons to 
the meson theory of the nucleon-nucleon interaction.
However, to avoid the problems with multi-meson exchanges and higher order corrections encountered during Phase~I, the various mesons were now exchanged just singly (i.~e., in lowest order). In addition, one would multiply the meson-nucleon vertices with form factors (``cutoffs'') to remove the singularities at short distances.
Clearly, this is not QFT anymore. It is a model motivated by the meson-exchange idea.
These models became known as one-boson-exchange (OBE) models, which were started in the early 1960's\footnote{The research devoted to the $NN$ interaction during the 1960's
has been thoroughly reviewed by Moravcsik~\cite{Mor72}.} 
and turned out to be very successful in terms of phenomenology.
Their popularity extended all the way into the 1990's.

\subsection{The One-Boson-Exchange Model}

\begin{figure}[t]\centering
%\vspace*{-0.25cm}
\scalebox{0.40}{\includegraphics{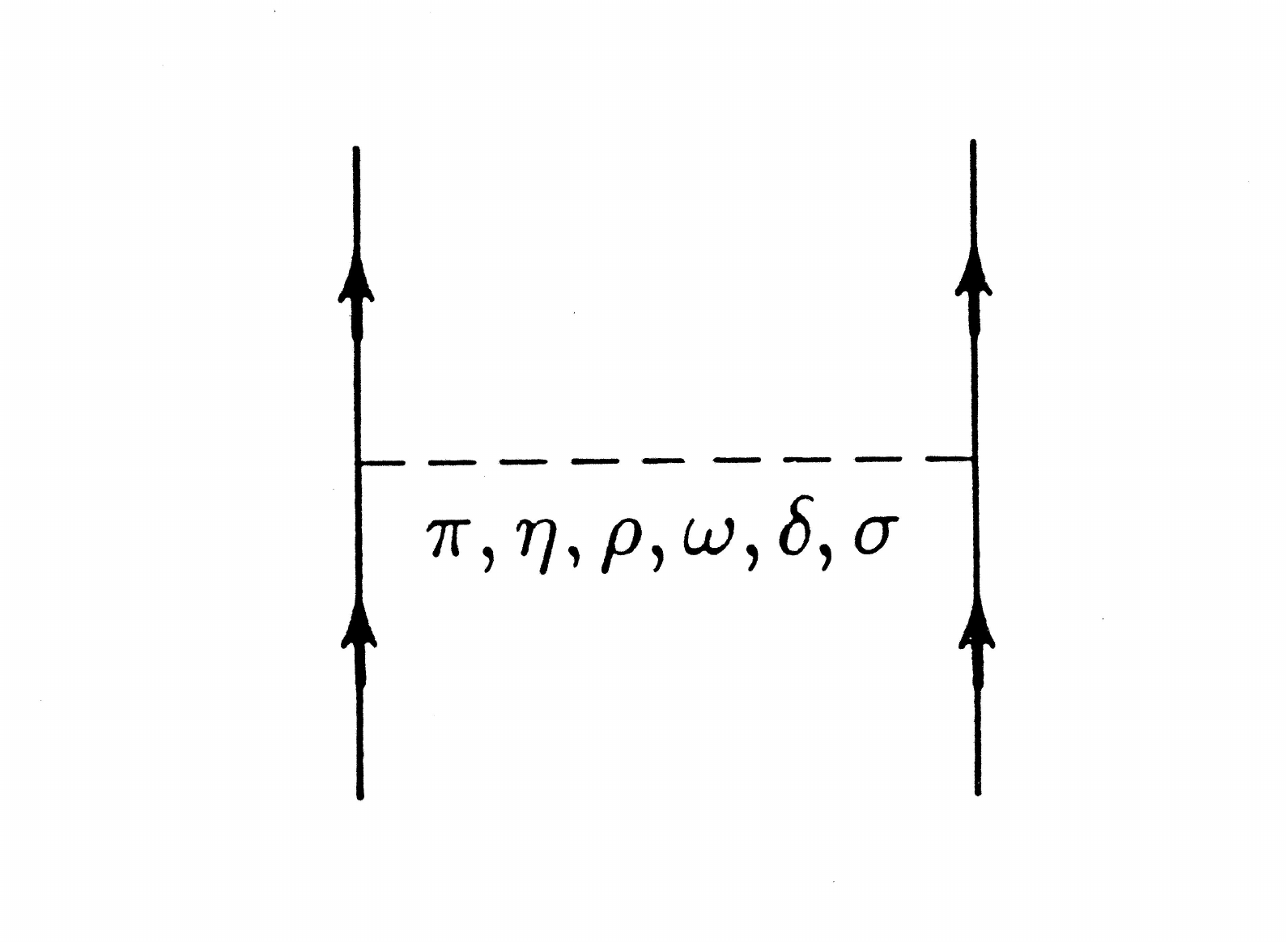}}
\vspace*{-0.5cm}
\caption{The one-boson-exchange model.
Solid lines denote nucleons and the dashed line represents mesons.}
\label{fig_obep}
\end{figure}

A typical one-boson-exchange model includes,
about half a dozen of bosons with masses up to about 1 GeV, Fig.~\ref{fig_obep}.
Not all mesons are equally important. The 
leading actors are the following four particles: 
\begin{itemize}
\item The pseudoscalar pion with a mass of about 138 MeV and isospin $I=1$ (isovector).
It is the lightest meson and provides the long-range part of the potential 
and most of the tensor force.
\item The isovector $\rho$ meson, a 2$\pi$ $P$-wave resonance of about 770 MeV.
Its major effect is to cut down the 
tensor force provided by the pion at short range.
\item The isoscalar $\omega$ meson, a 3$\pi$ resonance of 783 MeV and spin 1.
It creates a strong repulsive central 
force of short range (`repulsive core') and the nuclear spin-orbit force.
\item The scalar-isoscalar $f_0(500)$ or $\sigma$ boson with a mass around 500 MeV. It
 provides the crucial intermediate 
range attraction necessary for nuclear binding.
The interpretation as a particle is controversial~\cite{PDG}.
It may also be viewed  
as a simulation of effects of correlated $S$-wave 2$\pi$-exchange.
\end{itemize}
Obviously,
just these four
mesons can produce the major properties of 
the nuclear force.\footnote{The
interested reader can find a pedagogical
introduction into the OBE model in
sections~3 and 4 of Ref.~\cite{Mac89}.}
 
 Classic examples for OBE potentials (OBEPs) are the 
Bryan-Scott potentials started in the early 1960's~\cite{BS64}, but soon many other researchers got involved, too~\cite{Sup67,NRS78}. 
Since it is suggestive to think of a potential as a function of $r$
(where $r$ denotes the distance between the centers of the two interacting
nucleons), the OBEPs of the 1960's where represented as 
local $r$-space potentials. Some groups continued to hold on to this tradition and, thus, 
the construction of improved $r$-space OBEPs continued well into the 1990's~\cite{Sto94}.

An important advance during the 1970's  was the development of the
{\it relativistic OBEP}~\cite{ROBEP,Erk74,HM75}. In this model, the full, relativistic
Feynman amplitudes for the various one-boson-exchanges
are used to define the potential. These nonlocal expressions do not
pose any numerical problems  when used in momentum space and allow for a more quantitave descripton of $NN$ scattering.
The the high-precision CD-Bonn potential~\cite{Mac01}
is of this nature.

\subsection{Beyond the OBE approximation}
\label{sec_2pi}

\begin{figure}[t]\centering
%\vspace*{-0.25cm}
\scalebox{0.40}{\includegraphics{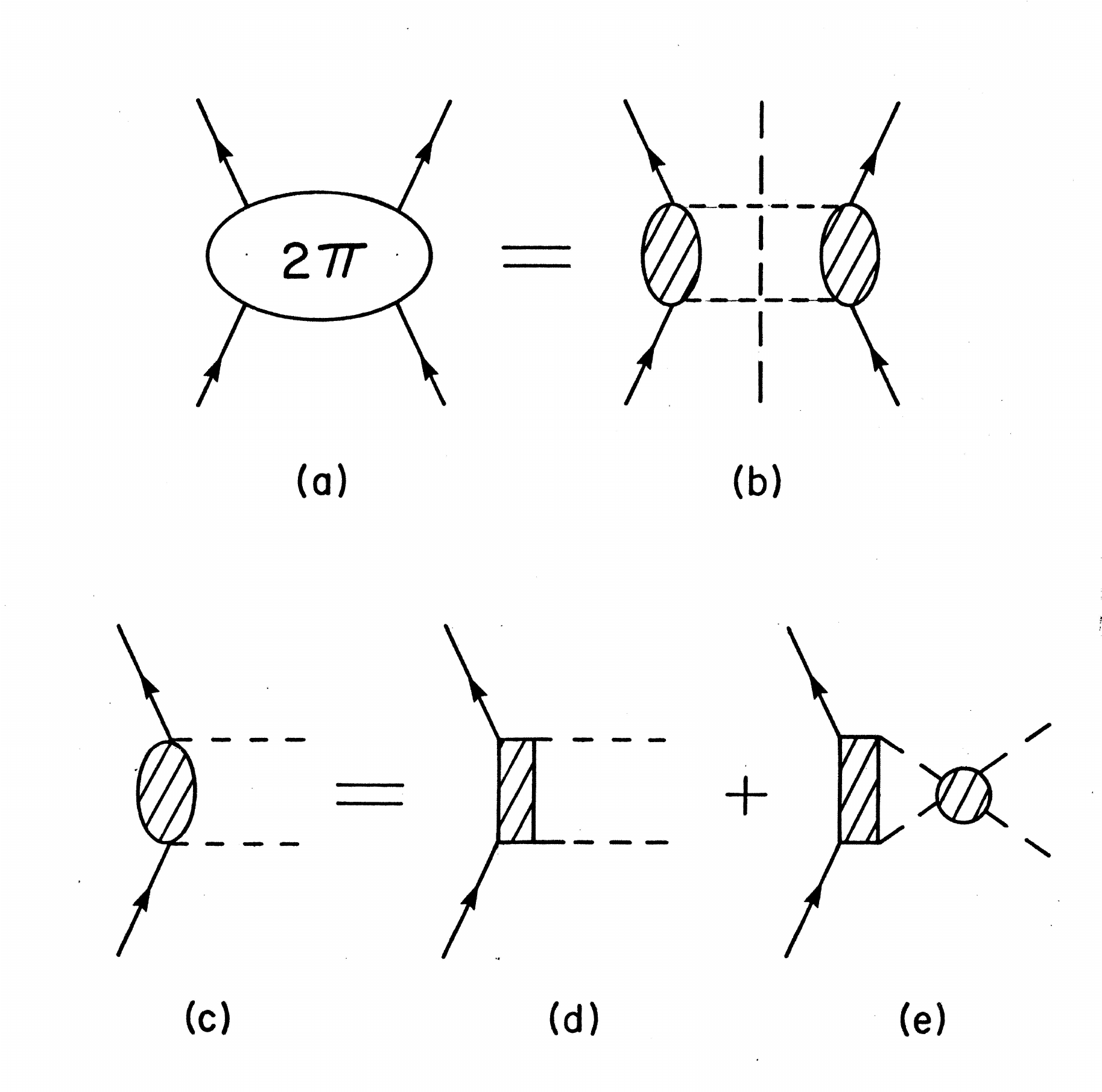}}
\vspace*{-0.2cm}
\caption{The $2\pi$-exchange contribution to the $NN$ interaction as viewed
by dispersion theory. Solid lines represent nucleons and dashed lines pions.
Further explanations are given in the text.}
\label{fig_disp}
\end{figure}

Historically, one must understand that,
after the failure of the pion theories in the 1950's,
the OBE model was considered a great
success in the 1960's~\cite{Sup67}.

On the other hand, one has to concede that
the OBE model is a great simplification of the complicated scenario of a full meson theory for the $NN$ interaction. Therefore, in spite of the
quantitative success of the OBEPs,
one should be concerned about the approximations involved in the model.
Major critical points include:
\begin{itemize}
\item
The scalar isoscalar $\sigma$ 'meson' of about 500 MeV.
\item
The neglect of all non-iterative diagrams.
\item
The role of meson-nucleon resonances.
\end{itemize}

Two pions, when 'in the air', can interact strongly. When in a relative $P$-wave $(L=1)$, they form a proper resonance,
the $\rho$ meson. They can also interact in a relative $S$-wave $(L=0)$, which gives rise to the $\sigma$ boson.
Whether the $\sigma$ is a proper resonance is controversial, even though the Particle Data Group
lists an $f_0(500)$ or $\sigma(500)$ meson, but with a width 400-700 MeV~\cite{PDG}. What is for sure is that two pions have correlations,
and if one doesn't believe in the $\sigma$ as a two pion resonance, then one has to take these correlations into account.
There are essentially two approaches that have been used to calculate these two-pion exchange contributions
to the $NN$ interaction (which generates the intermediate range attraction): dispersion theory and field theory.

In the 1960's, dispersion theory was developed out of frustration with the failure
of a QFT for strong interactions in the 1950's~\cite{Mor63}.
In the dispersion-theoretic approach the $NN$ amplitude is connected to the (empirical) $\pi N$ amplitude
by causality (analyticity), unitarity, and crossing symmetry.
Schematically this is shown in 
Fig.~\ref{fig_disp}.
The total diagram (a)
is analysed in terms of two 'halves' (b). The hatched ovals
stand for all possible  processes
which a pion and a nucleon
can undergo. This is made more explicit in (d) and (e).
The hatched boxes represent 
baryon intermediate states including the nucleon.
(Note that there are also
crossed pion exchanges which are not shown.)
The shaded circle
stands for $\pi \pi$ scattering.
Quantitatively, these processes are taken into account
by using  empirical
information from $\pi N$ and $\pi \pi$ scattering
(e. g. phase shifts) which 
represents the input for such a calculation.
Dispersion relations then provide an on-shell $NN$ amplitude, which
--- with some kind of plausible prescription --- is represented as
a potential.
The Stony Brook~\cite{JRV75,BJ76} and Paris~\cite{CV63,Vin73} 
groups have pursued this approach.
They could show that the intermediate-range part of the nuclear force
is, indeed, decribed about right by the $2 \pi$-exchange 
as obtained from dispersion
integrals.
To construct a complete potential, the $2\pi$-exchange contribution is
complemented by one-pion
and $\omega$ exchange. In addition to this, the Paris potential~\cite{Lac80}
 contains a 
phenomenological short-range part for $r< 1.5$ fm to improve the fit to the $NN$ data.
For further details,
we refer the interested reader to a pedagogical article by Vinh Mau~\cite{Vin79}.

\begin{figure}[t]\centering
%\vspace*{-0.25cm}
\scalebox{0.5}{\includegraphics{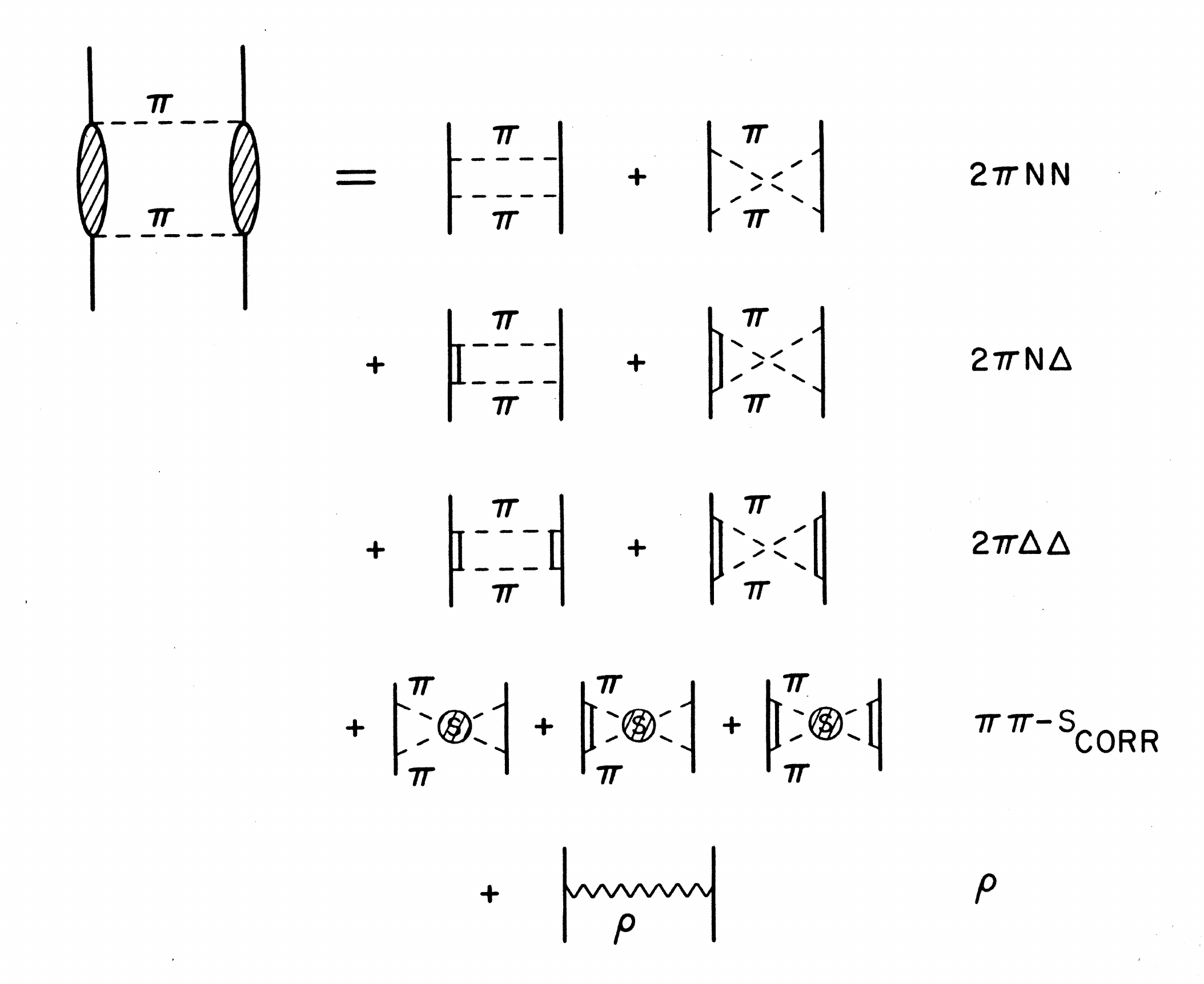}}
\vspace*{-0.2cm}
\caption{Field-theoretic model for the $2\pi$-exchange.
Notation as in Fig.~\ref{fig_disp}.
Double lines represent isobars.
The hatched circles are $\pi\pi$ correlations.
Further explanations are given in the text.}
\label{fig_2pi}
\end{figure}

A first field-theoretic attempt towards the $2\pi$-exchange was undertaken by Lomon and Partovi~\cite{PL70}.
Later, the more elaborated model shown in Fig.~\ref{fig_2pi} was developed by the Bonn group~\cite{MHE87}.
The model includes contributions from isobars as well as
from  $\pi \pi$ correlations.
This  can be understood
in analogy to the dispersion relations picture.
In general, only the lowest $\pi N$
resonance, the
so-called $\Delta$ isobar (spin 3/2, isospin 3/2, mass 1232 MeV),
is taken into account. The contributions from
other resonances have proven to be small for the low-energy $NN$
processes under consideration.
A field-theoretic model treats the
$\Delta$ isobar  as an elementary (Rarita-Schwinger) particle.
The six upper diagrams of Fig.~\ref{fig_2pi}
represent uncorrelated $2\pi$
exchange. The crossed (non-iterative)
two-particle exchanges
(second diagram in each row) are important.
They guarantee the proper (very weak) isospin dependence
due to characteristic cancelations in the isospin dependent
parts of box and crossed box diagrams.
Furthermore, their contribution is about as large as the one from
the corresponding box diagrams (iterative
diagrams); therefore, they are not negligible.
In addition to the processes discussed, also correlated $2\pi$
exchange has to be included (lower two rows of 
Fig.~\ref{fig_2pi}).
Quantitatively,
these contributions are about as sizable as those from the uncorrelated
processes.
Graphs with virtual pairs are left out, because
the pseudovector (gradient) coupling is used for the pion, in which case pair terms are small.

\begin{figure}[t]\centering
%\vspace*{-0.25cm}
\scalebox{0.30}{\includegraphics{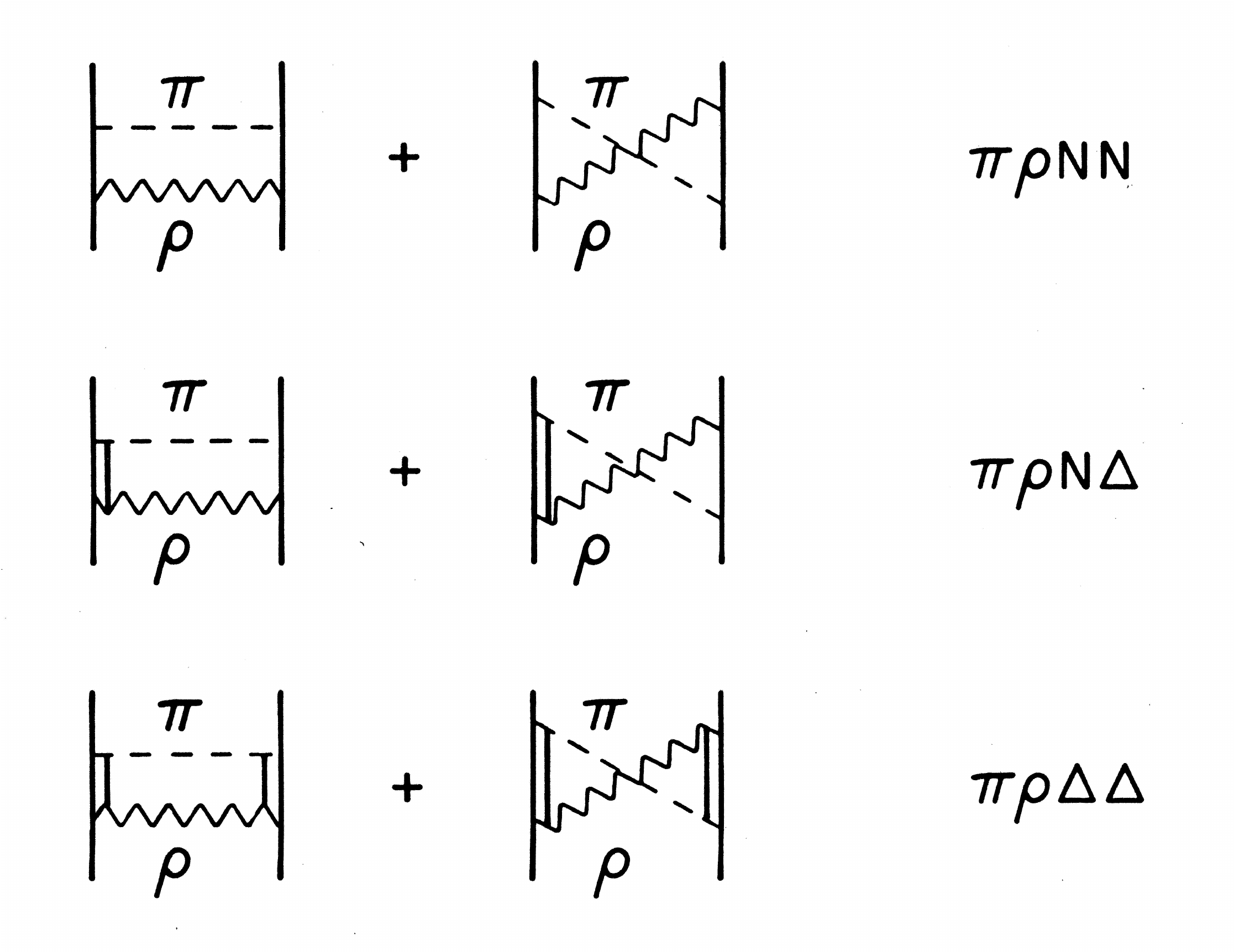}}
\vspace*{-0.2cm}
\caption{$\pi\rho$ contributions to the $NN$ interaction.}
\label{fig_pirho}
\end{figure}

Besides the contributions from two pions, there are also contributions
from the combination of other mesons. The combination of $\pi$ and $\rho$
is particularly significant, Fig.~\ref{fig_pirho}.
This contribution is repulsive and important
to suppress the 2$\pi$ exchange contribution at high momenta (or small distances), which is too strong by itself.

The Bonn Full Model~\cite{MHE87},
includes all the diagrams displayed in
Figs.~\ref{fig_2pi} and \ref{fig_pirho}
plus single $\pi$ and $\omega$ exchange.

\begin{figure}[t]\centering
\vspace*{-0.5cm}
\scalebox{0.6}{\includegraphics{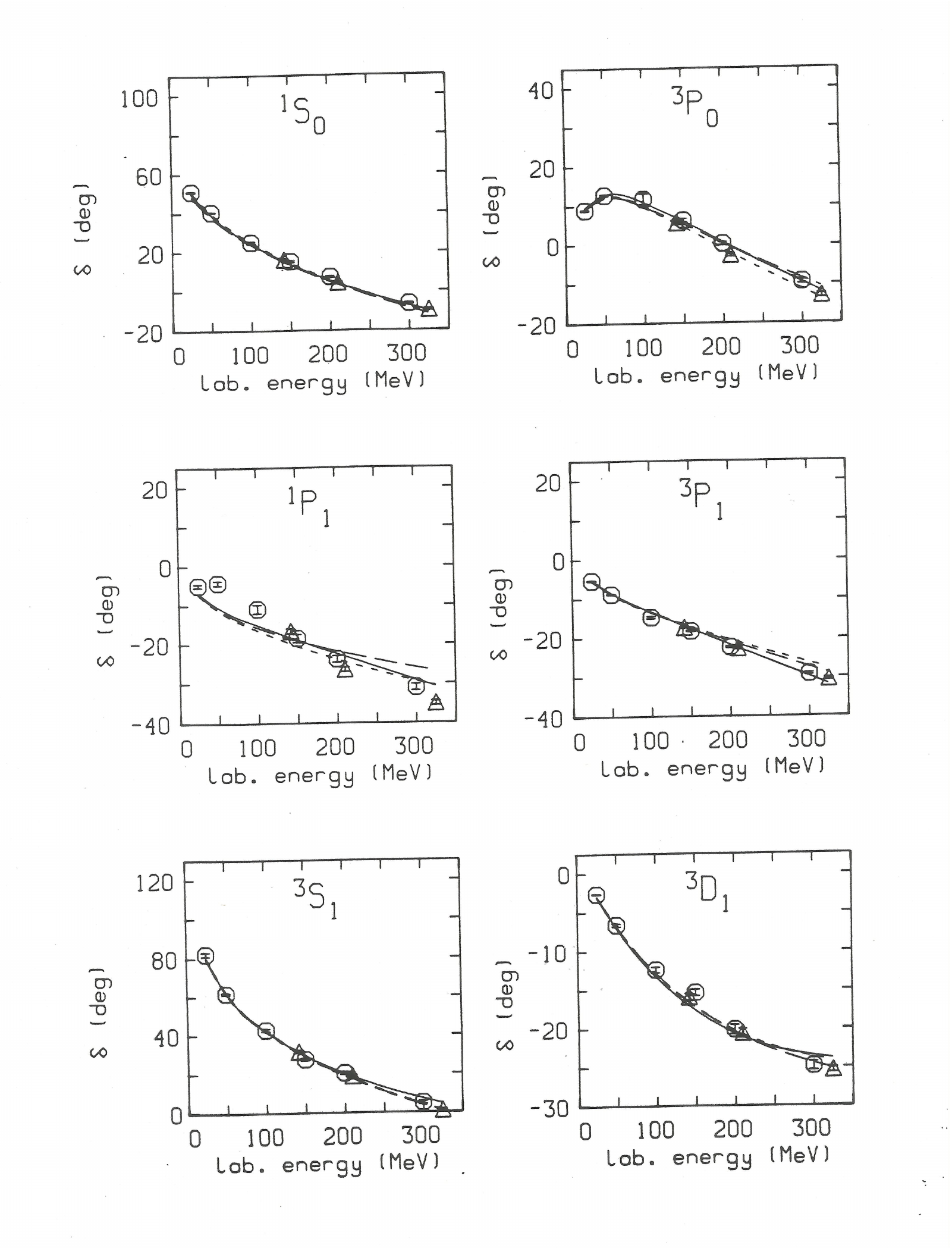}}
\vspace*{-0.5cm}
\caption{Phase shifts of $np$ scattering from some ``classic'' meson-exchange models for the $NN$ interaction. Predictions are shown for the Bonn Full Model~\cite{MHE87} (solid line), the Paris potential~\cite{Lac80} (dashed), and a (relativistic) OBEP~\cite{Mac89} (dotted). 
Phase parameters with total angular momentum $J\leq 1$ are displayed.
Symbols represent results from $NN$ phase shift analyses.}
\label{fig_phmeson1}
\end{figure}

\begin{figure}[t]\centering
\vspace*{-0.5cm}
\scalebox{0.6}{\includegraphics{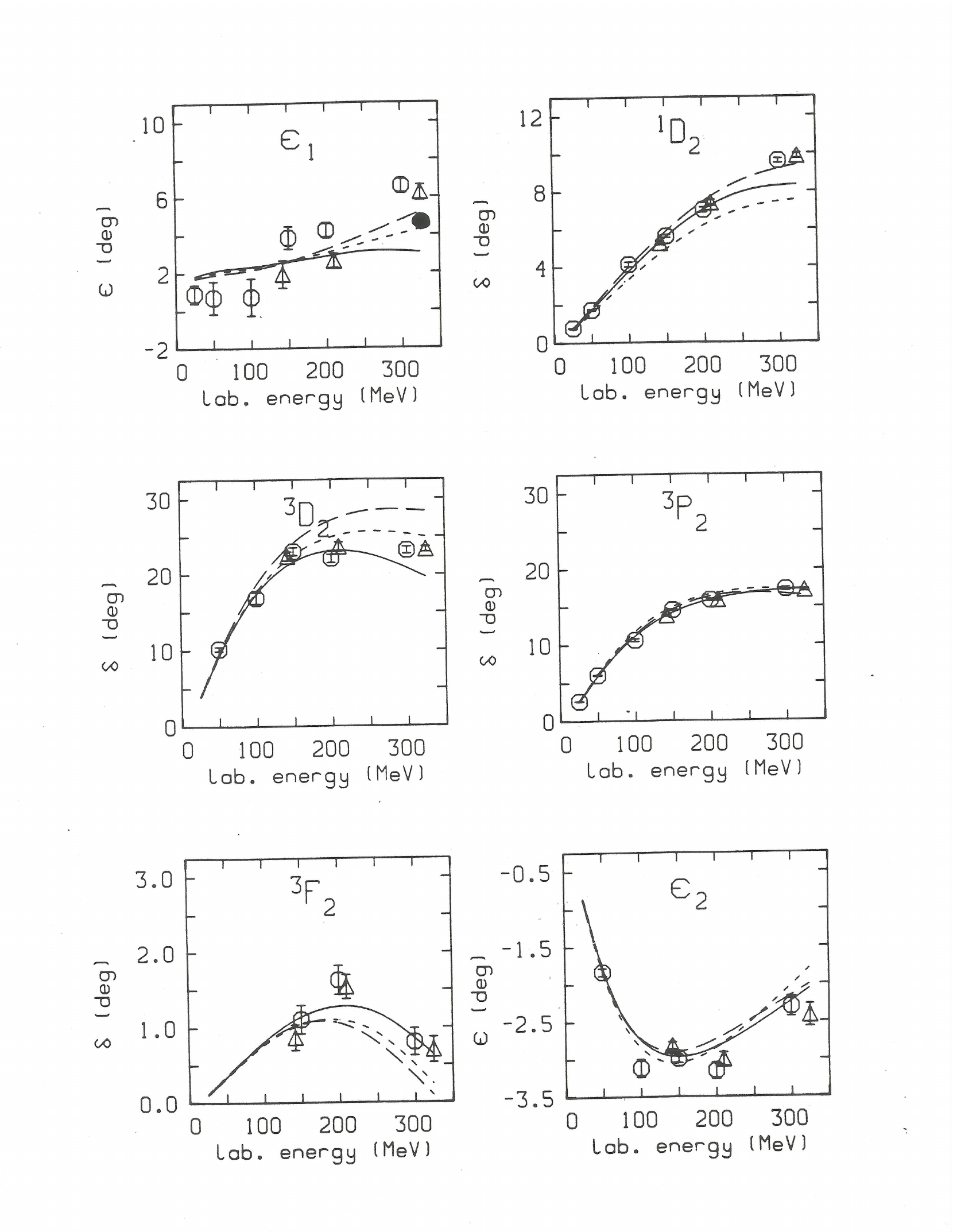}}
\vspace*{-0.5cm}
\caption{Same as Fig.~\ref{fig_phmeson1}, but $J=2$
phase shifts and $J\leq 2$ mixing parameters are shown.}
\label{fig_phmeson2}
\end{figure}

Having highly sophisticated models at hand, like the Paris and the Bonn potentials, allows to check
the approximations made in the simple OBE model. As it turns out, the complicated 2$\pi$ exchange
contributions to the $NN$ interaction tamed by the $\pi\rho$ diagrams can well be simulated by the single scalar isoscalar boson, 
the $\sigma$, with a mass around 550 MeV. In retrospect, this fact provides justification for the simple OBE model.
To illustate this point, we show, in Figs.~\ref{fig_phmeson1} and \ref{fig_phmeson2},
the phase shift predictions from the Bonn~\cite{MHE87} and Paris~\cite{Lac80} potentials as well as a relativistic OBEP~\cite{Mac89}.

The most important result of Phase II is that
meson exchange is an excellent phenomenology for describing nuclear forces.
It allows for the construction of very quantitative models. Therefore, 
the high-precision $NN$ potentials constructed in the mid-1990's are all
based upon meson phenomenology~\cite{Sto94,Mac01,WSS95}.
However, with the rise of QCD to
the ranks of the authoritative theory of strong interactions, meson-exchange is definitively just a model. This implies that
if we ultimately wish to solve the nuclear force problem on the most fundamental grounds,
then we have to begin all over again---starting with QCD.

\section{Phase III (1990 -- 2020): Chiral Effective Field Theory}
\label{sec_cheft}

\subsection{QCD and nuclear forces}

Quantum chromodynamics provides the theoretical framework
to describe strong interactions, namely interactions involving  
quarks and gluons. According to QCD, objects which carry color
interact weakly at short distances and strongly at large
distances, where the separation between the two regimes is
about 1 fm. Naturally, short distances and long distances can be
associated with high and low energies, respectively, causing the     
quarks to be confined into hadrons, which carry no color. 
At the same time, the weak nature of the force at high energies 
results into what is known as ``asymptotic freedom". 
(We note that these behaviors originate from the fact that 
QCD is a non-Abelian gauge field theory
with color $SU(3)$ the underlying gauge group.) 
Therefore, QCD is perturbative at high energy, 
but strongly coupled at low-energy.             
The energies typical for nuclear physics are low and, thus,
nucleons are appropriate degrees of freedom. The nuclear force 
can then be regarded as a      
residual color interaction acting between nucleons in a way similar to how
the van der Waals forces bind neutral molecules.
If described in terms of quark and gluon degrees of freedom, the interaction  
between nucleons is an extremely complex problem.

Therefore, during the first round of new attempts, QCD-inspired quark 
models~\cite{EFV00} 
became popular. 
The positive aspect of these models is that they try to explain hadron structure
and hadron-hadron interactions on an equal footing and, indeed, 
some of the gross features of the $NN$ interaction are explained successfully.
However, on a critical note, it must be pointed out
that these quark-based
approaches are nothing but
another set of models and, thus, do not represent
fundamental progress. 
For the purpose of describing hadron-hadron interactions, 
one may equally well then stay
with the simpler and much more quantitative meson models.

Alternatively, one may try to solve the problem with brute computing power by a
method known as lattice QCD.
In a recent paper~\cite{Org15}, the 
nucleon-nucleon system is investigated at a pion mass of about 450 MeV.
Over the range of energies that are studied, the scattering phase shifts in the $^1S_0$ and 
$^3S-D_1$ channels are found to be similar to those in nature and indicate a repulsive
short-range component of the interaction.
 This result is then extrapolated
to the physical pion mass with the help of chiral perturbation theory. The pion mass
of 450 MeV is still too large to allow for reliable extrapolations, but the feasibility has been
demonstrated and more progress can be expected for the near future.
In a lattice calculation of a very different kind, the $NN$ potential
was studied in Ref.~\cite{Hat12}. The central component of this potential exhibits 
repulsion at the core as well as intermediate-range attraction. This is encouraging, but 
one must keep in mind that the pion masses employed in this study are still quite large.
In summary, although calculations within lattice QCD are being performed and improved, 
 they are computationally very costly, and thus they are useful, in practice, only to explore a few cases.  
 Clearly, a different approach is necessary to address a full variety of nuclear structure problems.

\subsection{An EFT for low-energy QCD}
\label{sec_eft}

Around 1980/1990, a major breakthrough occurred when the nobel laureate Steven Weinberg applied the concept of an effective field theory (EFT) to low-energy QCD~\cite{Wei79,GL84,GSS88,Wei90,Wei91}. He simply wrote down the most general Lagrangian that is consistent with all the properties of low-energy QCD, since that would make this theory equivalent to low-energy QCD. A particularly important property is the so-called chiral symmetry, which is ``spontaneously'' broken. 
 The effective degrees of freedom are then pions 
(the Goldstone bosons of the broken symmetry) and
nucleons rather than quarks and gluons; heavy mesons and
nucleon resonances are ``integrated out''.
So, the circle of history is closing and we are {\it back to a pion theory} (cf.\ Phase I)
except that we have finally learned how to deal with it:
broken chiral symmetry is a crucial constraint that generates
and controls the dynamics and establishes a clear connection
with the underlying theory, QCD.
The constraint of chiral symmetry dictates that the pion couples to the nucleon via a
derivative coupling ($\gamma_5 \gamma^\mu \partial_\mu$).
Recall that this coupling was already considered in the 1950's~\cite{SBH55},
but discarded because it is not renormalizable.
In the context of a fundamental quantum field theory, this coupling is, indeed, not renormalizable.
However, the scenario is now different~\cite{Wei97,Wei09}.
The gradient coupling is revived in the context of an {\it effective} field theory.
Such a theory, is organized order by order and, therefore, renormalized order by order. In each order,
only a finite number of counter terms is needed to renormalize. Moreover, the calculation is carried out only up to a finite order at which the desired accuracy is achieved. Thus, everything is manageable.

For the order by order expansion of the EFT, an appropriate ``large scale'' needs to be identified.
The large difference between the masses of
the pions and the masses of the vector mesons, like $\rho(770)$ and $\omega(782)$, provides a clue.
From that observation, one is prompted to take the pion mass as the identifier of the soft scale, 
$Q \sim m_\pi$,
while the rho mass sets the hard scale, $\Lambda_\chi \sim m_\rho$, often referred to 
as the chiral-symmetry breaking scale.
It is then natural to 
consider an expansion in terms of                                            
$Q/\Lambda_\chi$.

To summarize, an EFT program for nuclear forces involves the following steps:                      
\begin{enumerate}
\item
Identify the low- and high-energy scales, and the degrees of freedom suitable       
for (low-energy) nuclear physics.
\item
Recognize the symmetries of low-energy QCD and
explore the mechanisms responsible of their breakings.                         
\item
Build the most general Lagrangian which respects those
(broken) symmetries.                        
\item
Formulate a scheme to organize contributions in order 
of their importance. 
Clearly, this amounts to performing an expansion in 
terms of (low) momenta. 
\item
Using the expansion mentioned above, evaluate Feynman diagrams
to the desired accuracy.
\end{enumerate}

In what follows, we will discuss each of the steps above. 
Note that the first one has already been addressed, so 
we will move directly to the second one.

\subsection{Symmetries of low-energy QCD}

Our purpose here is to provide a compact introduction into 
(low-energy) QCD, with particular attention to the symmetries
and their breakings. 
For more details the reader is referred to                 
Refs.~\cite{ME11,Sch03}.

\subsubsection{Chiral symmetry}

We begin with the QCD Lagrangian,         
\begin{equation}
{\cal L}_{\rm QCD} = 
\bar{q} (i \gamma^\mu {\cal D}_\mu - {\cal M})q
 - \frac14 
{\cal G}_{\mu\nu,a}
{\cal G}^{\mu\nu}_{a} 
\label{eq_LQCD}
\end{equation}
with the gauge-covariant derivative
\begin{equation}
{\cal D}_\mu = \partial_\mu -ig\frac{\lambda_a}{2}
{\cal A}_{\mu,a}
\label{eq_Dm}
\end{equation}
and the gluon field strength tensor\footnote{For $SU(N)$ group indices, we use 
Latin letters, $\ldots,a,b,c,\ldots,i,j,k,\dots$,
and, in general, do not distinguish between subscripts and superscripts.}
\begin{equation}
{\cal G}_{\mu\nu,a} =
\partial_\mu {\cal A}_{\nu,a}
-\partial_\nu {\cal A}_{\mu,a}
 + g f_{abc}
{\cal A}_{\mu,b}
{\cal A}_{\nu,c} \,.
\label{eq_Gmn}
\end{equation}
In the above, $q$ denotes the quark fields and ${\cal M}$
the quark mass matrix. Further, $g$ is the
strong coupling constant and ${\cal A}_{\mu,a}$
are the gluon fields. 
Moreover,
$\lambda_a$ are the
Gell-Mann matrices and $f_{abc}$
the structure constants of the $SU(3)_{\rm color}$
Lie algebra $(a,b,c=1,\dots ,8)$;
summation over repeated indices is always implied.
The gluon-gluon term in the last equation arises
from the non-Abelian nature of the gauge theory
and is the reason for the peculiar features
of the color force.

The current masses of the up $(u)$, down $(d)$, and
strange (s) quarks are in a $\overline{MS}$ scheme at a scale of $\mu \approx 2$ GeV~\cite{PDG}:
\begin{eqnarray}
m_u &=& 2.3\pm 0.7 \mbox{ MeV} ,
\label{eq_umass} \\
m_d &=& 4.8\pm 0.5 \mbox{ MeV} ,
\label{eq_dmass} \\
m_s &=& 95\pm 5 \mbox{ MeV} .
\label{eq_smass}
\end{eqnarray}
These masses are small as compared to
a typical hadronic scale
such as the mass of a light hadron other than a Goldstone bosons, e.g., 
$m_\rho=0.78 \mbox{ GeV} \approx 1 \mbox{ GeV}$.

Thus it is relevant to discuss the
QCD Lagrangian in the case when the quark masses
vanish:  
\begin{equation}
{\cal L}_{\rm QCD}^0 = \bar{q} i \gamma^\mu {\cal D}_\mu
q - \frac14 
{\cal G}_{\mu\nu,a}
{\cal G}^{\mu\nu}_{a} \,.
\end{equation}
Right- and left-handed quark fields are defined as
\begin{equation}
q_R=P_Rq \,, \;\;\;
q_L=P_Lq \,,
\end{equation}
with 
\begin{equation}
P_R=\frac12(1+\gamma_5) \,, \;\;\;
P_L=\frac12(1-\gamma_5) \,.
\end{equation}
Then the Lagrangian can be rewritten as 
\begin{equation}
{\cal L}_{\rm QCD}^0 = 
\bar{q}_R i \gamma^\mu {\cal D}_\mu q_R 
+\bar{q}_L i \gamma^\mu {\cal D}_\mu q_L 
- \frac14 
{\cal G}_{\mu\nu,a}
{\cal G}^{\mu\nu}_{a} \, .
\end{equation}
This equation revels that
{\it the right- and left-handed components of
massless quarks do not mix} in the QCD Lagrangian. 
For the two-flavor case, this is
$SU(2)_R\times SU(2)_L$ 
symmetry, also known as {\it chiral symmetry}.
However, this symmetry is broken in two ways: explicitly and spontaneously.

\subsubsection{Explicit symmetry breaking}

The mass term  
 $- \bar{q}{\cal M}q$
in the QCD Lagrangian Eq.~(\ref{eq_LQCD}) 
breaks chiral symmetry explicitly. To better see this,
let's rewrite ${\cal M}$ for the two-flavor case,
\begin{eqnarray}
{\cal M} & = & 
\left( \begin{array}{cc}
            m_u & 0 \\
              0  & m_d 
           \end{array} \right)  \nonumber \\
  & = & \frac12 (m_u+m_d) 
\left( \begin{array}{cc}
            1 & 0 \\
              0  & 1 
           \end{array} \right) 
+ \frac12 (m_u-m_d) 
\left( \begin{array}{cc}
            1 & 0 \\
              0  & -1 
           \end{array} \right)  \nonumber \\
 & = & \frac12 (m_u+m_d) \; I + \frac12 (m_u-m_d) \; \tau_3 \,.
\label{eq_mmatr}
\end{eqnarray}
The first term in the last equation in invariant under $SU(2)_V$
(isospin symmetry) and the second term vanishes for
$m_u=m_d$.
Therefore, isospin is an exact symmetry if 
$m_u=m_d$.
However, both terms in Eq.~(\ref{eq_mmatr}) break chiral symmetry.
Since the up and down quark masses
[Eqs.~(\ref{eq_umass}) and (\ref{eq_dmass})]
are small as compared to
the typical hadronic mass scale of $\sim 1$ GeV,
the explicit chiral symmetry breaking due to non-vanishing
quark masses is very small.

\subsubsection{Spontaneous symmetry breaking}

A (continuous) symmetry is said to be {\it spontaneously
broken} if a symmetry of the Lagrangian 
is not realized in the ground state of the system.
There is evidence that the (approximate) chiral
symmetry of the QCD Lagrangian is spontaneously 
broken---for dynamical reasons of nonperturbative origin
which are not fully understood at this time.
The most plausible evidence comes from the hadron spectrum.

From chiral symmetry,
one naively expects the existence of degenerate hadron
multiplets of opposite parity, i.e., for any hadron of positive
parity one would expect a degenerate hadron state of negative 
parity and vice versa. However, these ``parity doublets'' are
not observed in nature. For example, take the $\rho$-meson which is
a vector meson of negative parity ($J^P=1^-$) and mass 
776 MeV. There does exist a $1^+$ meson, the $a_1$, but it
has a mass of 1230 MeV and, therefore, cannot be perceived
as degenerate with the $\rho$. On the other hand, the $\rho$
meson comes in three charge states (equivalent to
three isospin states), the $\rho^\pm$ and the $\rho^0$,
with masses that differ by at most a few MeV. Thus,
in the hadron spectrum,
$SU(2)_V$ (isospin) symmetry is well observed,
while axial symmetry is broken:
$SU(2)_R\times SU(2)_L$ is broken down to $SU(2)_V$.

A spontaneously broken global symmetry implies the existence
of (massless) Goldstone bosons.
The Goldstone bosons are identified with the isospin
triplet of the (pseudoscalar) pions, 
which explains why pions are so light.
The pion masses are not exactly zero because the up
and down quark masses
are not exactly zero either (explicit symmetry breaking).
Thus, pions are a truly remarkable species:
they reflect spontaneous as well as explicit symmetry
breaking.
Goldstone bosons interact weakly at low energy.
They are degenerate with the vacuum 
and, therefore, interactions between them must
vanish at zero momentum and in the chiral limit
($m_\pi \rightarrow 0$).

\subsection{Chiral effective Lagrangians 
\label{sec_Lpi} }

The next step in our EFT program is to build the most general
Lagrangian consistent with the (broken) symmetries discussed
above.
An elegant formalism for the construction of such Lagrangians
was developed by 
Callan, Coleman, Wess, and Zumino (CCWZ)~\cite{CCWZ}
who developed the foundations 
of non-linear realizations of chiral symmetry from the point of view
of group theory.\footnote{An accessible introduction into the rather
involved CCWZ formalism can be found in Ref.~\cite{Sch03}.}
The Lagrangians we give below are built upon the CCWZ
formalism. 

We already addressed the fact that the appropriate degrees of freedom are
pions (Goldstone bosons) and nucleons.
Because pion interactions must      
vanish at zero momentum transfer and in the limit of    
$m_\pi \rightarrow 0$, namely the chiral limit, the                         
Lagrangian is expanded in powers of derivatives
and pion masses. More precisely, the Lagrangian is expanded in powers of 
$Q/\Lambda_\chi$ where $Q$ stands for a (small) momentum
or pion mass and 
$\Lambda_\chi \approx 1$ GeV is identified with the              
hard scale.                                  
These are the basic steps behind the chiral perturbative expansion.    

Schematically, we can write the effective Lagrangian as                   
\begin{equation}
{\cal L}
=
{\cal L}_{\pi\pi} 
+
{\cal L}_{\pi N} 
+
{\cal L}_{NN} 
 + \, \ldots \,,
\end{equation}
where ${\cal L}_{\pi\pi}$
deals with the dynamics among pions, 
${\cal L}_{\pi N}$ 
describes the interaction
between pions and a nucleon,
and ${\cal L}_{NN}$  contains two-nucleon contact interactions
which consist of four nucleon-fields (four nucleon legs) and no
meson fields.
The ellipsis stands for terms that involve two nucleons plus
pions and three or more
nucleons with or without pions, relevant for nuclear
many-body forces 
(an example for this in lowest order are the last two terms of Eq.~(\ref{eq_LD1}), below).
The individual Lagrangians are organized order by order:
\begin{equation}
{\cal L}_{\pi\pi} 
 = 
{\cal L}_{\pi\pi}^{(2)} 
 + {\cal L}_{\pi\pi}^{(4)} 
 + \ldots \,,
\end{equation}
\begin{equation}
{\cal L}_{\pi N} 
= 
{\cal L}_{\pi N}^{(1)} 
+
{\cal L}_{\pi N}^{(2)} 
+
{\cal L}_{\pi N}^{(3)} 
+
{\cal L}_{\pi N}^{(4)} 
+
{\cal L}_{\pi N}^{(5)} 
+ \ldots ,
\end{equation}
and
\begin{equation}
\label{eq_LNN}
{\cal L}_{NN} =
{\cal L}^{(0)}_{NN} +
{\cal L}^{(2)}_{NN} +
{\cal L}^{(4)}_{NN} + 
{\cal L}^{(6)}_{NN} + 
\ldots \,,
\end{equation}
where the superscript refers to the number of derivatives or 
pion mass insertions (chiral dimension)
and the ellipsis stands for terms of higher dimensions.

Above, we have organized the Lagrangians by the number
of derivatives or pion-masses. This is 
the standard way, appropriate particularly for
considerations of $\pi$-$\pi$ and $\pi$-$N$ scattering.
As it turns out (cf.\ Section~\ref{sec_chpt}), 
for interactions among nucleons,
sometimes one makes use of the so-called
index of the interaction,
\begin{equation}
\Delta  \equiv   d + \frac{n}{2} - 2  \, ,
\label{eq_Delta}
\end{equation}
where $d$ is the number of derivatives or pion-mass insertions 
and $n$ the number of nucleon field operators (nucleon legs).
We will now write down the Lagrangian in terms
of increasing values of the parameter $\Delta$ and
we will do so using the so-called heavy-baryon formalism which we indicate
by a ``hat''~\cite{Fet00}.

The leading-order Lagrangian reads,
\begin{eqnarray}
\widehat{\cal L}^{\Delta=0} &=&
	\frac{1}{2} 
	\partial_\mu \mbox{\boldmath{$\pi$}} \cdot 
	\partial^\mu \mbox{\boldmath{$\pi$}}
     -  \frac{1}{2} m_\pi^2 \mbox{\boldmath{$\pi$}}^2
\nonumber \\ &&
     +  \frac{1-4\alpha}{2f_\pi^2} 
        (\mbox{\boldmath{$\pi$}} \cdot \partial_\mu \mbox{\boldmath{$\pi$}})
        (\mbox{\boldmath{$\pi$}} \cdot \partial^\mu \mbox{\boldmath{$\pi$}})
     -  \frac{\alpha}{f_\pi^2} 
        \mbox{\boldmath{$\pi$}}^2
	\partial_\mu \mbox{\boldmath{$\pi$}} \cdot 
	\partial^\mu \mbox{\boldmath{$\pi$}}
     +\;  \frac{8\alpha-1}{8f_\pi^2} 
        m_\pi^2 \mbox{\boldmath{$\pi$}}^4
\nonumber \\ &&
+ \bar{N} \left[ 
i \partial_0 
- \frac{g_A}{2f_\pi} \; \mbox{\boldmath $\tau$} \cdot 
 ( \vec \sigma \cdot \vec \nabla ) \mbox{\boldmath $\pi$} 
- \frac{1}{4f_\pi^2} \; \mbox{\boldmath $\tau$} \cdot 
 ( \mbox{\boldmath $\pi$}
\times \partial_0 \mbox{\boldmath $\pi$})
\right] N
\nonumber \\ &&
+ \bar{N} \left\{
 \frac{g_A(4\alpha-1)}{4f_\pi^3} \;
(\mbox{\boldmath $\tau$} \cdot 
\mbox{\boldmath $\pi$}) 
\left[ \mbox{\boldmath $\pi$} \cdot 
 ( \vec \sigma \cdot \vec \nabla )
\mbox{\boldmath $\pi$} \right]
 \right. \nonumber \\ &&   \left.
+ \; \frac{g_A\alpha}{2f_\pi^3} \;
\mbox{\boldmath $\pi$}^2 
\left[ \mbox{\boldmath $\tau$} \cdot 
 ( \vec \sigma \cdot \vec \nabla )
\mbox{\boldmath $\pi$} 
\right]
\right\} N 
\nonumber \\ &&
-\frac{1}{2} C_S \bar{N} N \bar{N} N 
-\frac{1}{2} C_T (\bar{N} \vec \sigma N) \cdot (\bar{N} \vec \sigma N) 
\; + \; \ldots \,,
\label{eq_LD0}
\end{eqnarray}
and subleading Lagrangians are,
\begin{eqnarray}
\widehat{\cal L}^{\Delta=1} &=&
 \bar{N} \left\{
 \frac{{\vec \nabla}^2}{2M_N} 
-\frac{ig_A}{4M_Nf_\pi} 
\mbox{\boldmath $\tau$} \cdot 
\left[
\vec \sigma \cdot
\left( \stackrel{\leftarrow}{\nabla} 
\partial_0 \mbox{\boldmath $\pi$}
 -
\partial_0 \mbox{\boldmath $\pi$}
\stackrel{\rightarrow}{\nabla} \right)
\right]
\right.
\nonumber \\ &&
\left.
- \frac{i}{8M_N f_\pi^2}
\mbox{\boldmath $\tau$} \cdot 
\left[
\stackrel{\leftarrow}{\nabla} 
\cdot
( \mbox{\boldmath $\pi$} \times \vec\nabla \mbox{\boldmath $\pi$} )
   -   
( \mbox{\boldmath $\pi$} \times \vec\nabla \mbox{\boldmath $\pi$} )
\cdot
\stackrel{\rightarrow}{\nabla} 
\right]
\right\} N 
\nonumber \\ &&
+ \bar{N} \left[
 4c_1m_\pi^2
-\frac{2 c_1}{f_\pi^2} \, m_\pi^2\, \mbox{\boldmath $\pi$}^2 
\, + \, 
\left( c_2 - \frac{g_A^2}{8M_N}\right) 
\frac{1}{f_\pi^2}
(\partial_0 \mbox{\boldmath{$\pi$}} \cdot 
 \partial_0 \mbox{\boldmath{$\pi$}})
\right.  \nonumber \\ &&  \left.
 + \, \frac{c_3}{f_\pi^2}\,
(\partial_\mu \mbox{\boldmath{$\pi$}} \cdot 
\partial^\mu \mbox{\boldmath{$\pi$}})
%\right.  \nonumber \\ &&  \left.
 - \, \left( c_4 + \frac{1}{4M_N} \right) 
\frac{1}{2f_\pi^2}
\epsilon^{ijk} \epsilon^{abc} \sigma^i \tau^a
(\partial^j \pi^b) (\partial^k \pi^c)
 \right] N 
\nonumber \\ &&
- \frac{D}{4f_\pi} (\bar{N}N) \bar{N} \left[ 
\mbox{\boldmath $\tau$} 
\cdot ( \vec \sigma \cdot \vec \nabla )
\mbox{\boldmath $\pi$} 
\right] N
\nn &&
-\frac12 E
(\bar{N}N)
(\bar{N}
\mbox{\boldmath $\tau$} 
N)
\cdot
(\bar{N}
\mbox{\boldmath $\tau$} 
N)
\; + \; \ldots \,,
\label{eq_LD1}
\\
\widehat{\cal L}^{\Delta=2} &=&
\; {\cal L}^{(4)}_{\pi\pi} \; +
\; \widehat{\cal L}^{(3)}_{\pi N} \; + \; \widehat{\cal L}^{(2)}_{NN}
\; + \; \ldots \,,
\label{eq_LD2}
\\
\widehat{\cal L}^{\Delta=3} &=&
\; \widehat{\cal L}^{(4)}_{\pi N} 
\; + \; \ldots \,,
\label{eq_LD3}
\\
\widehat{\cal L}^{\Delta=4} &=&
\; \widehat{\cal L}^{(5)}_{\pi N} \; +
\; \widehat{\cal L}^{(4)}_{NN}
\; + \; \ldots \,.
\label{eq_LD4}
\\
\widehat{\cal L}^{\Delta=6} &=&
\; \widehat{\cal L}^{(6)}_{NN}\; +
\;  \ldots \,,
\label{eq_LD6}
\end{eqnarray}
where we included terms relevant for a calculation of the two-nucleon force up to sixth order.
The Lagrangians $\widehat{\cal L}^{(3)}_{\pi N}$ and
$\widehat{\cal L}^{(4)}_{\pi N}$ 
can be found in Ref.~\cite{KGE12}.
The pion fields are denoted by $\boldsymbol{\pi}$ and the heavy baryon nucleon field by $N$  ($\bar{N}=N^\dagger$).
Furthermore, $g_A$, $f_\pi$, $m_\pi$, and $M_N$ are the axial-vector coupling constant,
pion decay constant, pion mass, and nucleon mass, respectively. 
The $c_i$ are low-energy constants (LECs) from the dimension two $\pi N$ Lagrangian 
and $\alpha$ is a parameter that appears in the expansion 
of a $SU(2)$ matrix $U$ in powers
of the pion fields, see Ref.~\cite{ME11} for more details. Results are independent of $\alpha$.

The lowest order (or leading order) $NN$ Lagrangian has no derivatives 
and reads~\cite{Wei91}
\begin{equation}
\label{eq_LNN0}
\widehat{\cal L}^{(0)}_{NN} =
-\frac{1}{2} C_S \bar{N} N \bar{N} N 
-\frac{1}{2} C_T (\bar{N} \vec \sigma N) \cdot (\bar{N} \vec \sigma N) \, ,
\end{equation}
where 
$C_S$ and $C_T$ are free paramters to be determined by 
fitting to the $NN$ data.

The second order $NN$ Lagrangian, $\widehat{\cal L}^{(2)}_{NN}$, can be found
in Ref.~\cite{ORK96}.
The $NN$ contact potentials derived from some of the $NN$ Lagrangians
are given in Sec.~\ref{sec_ct}.

\subsection{Nuclear forces from EFT: Overview
\label{sec_overview}}

We proceed here with discussing the various steps towards a derivation
of nuclear forces from EFT.
In this section, we will discuss the expansion
we are using in more details as well as the various 
Feynman diagrams as they emerge at each order. 

\subsubsection{Chiral perturbation theory and power counting
\label{sec_chpt}}

An infinite number of Feynman diagrams can be evaluated from 
the effective Langrangians and so one needs to be able to organize these
diagrams in order of their importance. 
Chiral perturbation theory (ChPT) provides such organizational scheme. 

In ChPT, 
graphs are analyzed
in terms of powers of small external momenta over the large scale:
$(Q/\Lambda_\chi)^\nu$,
where $Q$ is generic for a momentum (nucleon three-momentum or
pion four-momentum) or a pion mass and $\Lambda_\chi \sim 1$ GeV
is the chiral symmetry breaking scale (hadronic scale, hard scale).
Determining the power $\nu$ 
has become known as power counting.

For the moment, we will consider only so-called irreducible
graphs.
By definition, an irreducible graph is a diagram that
cannot be separated into two
by cutting only nucleon lines.
Following the Feynman rules of covariant perturbation theory,
a nucleon propagator carries the dimension $Q^{-1}$,
a pion propagator $Q^{-2}$,
each derivative in any interaction is $Q$,
and each four-momentum integration $Q^4$.
This is also known as naive dimensional analysis.
Applying then some topological identities, one obtains
for the power of an irreducible diagram
involving $A$ nucleons~\cite{ME11}
\begin{equation} \nu = -2 +2A - 2C + 2L 
+ \sum_i \Delta_i \, ,
\label{eq_nu} 
\end{equation}
with
\begin{equation}
\Delta_i  \equiv   d_i + \frac{n_i}{2} - 2  \, .
\label{eq_Deltai}
\end{equation}
In the two equations above: for each
vertex $i$, $C$ represents the number of individually connected parts of the diagram while
$L$ is the number of loops;                  
$d_i$ indicates how many derivatives or pion masses are present 
and $n_i$ the number of nucleon fields.                  
The summation extends over all vertices present in that particular diagram.
Notice also that chiral symmetry implies $\Delta_i \geq 0$. 
Interactions among pions have at least two derivatives
($d_i\geq 2, n_i=0$), while 
interactions between pions and a nucleon have one or more 
derivatives  
($d_i\geq 1, n_i=2$). Finally, pure contact interactions
among nucleons ($n_i=4$)
have $d_i\geq0$.
In this way, a low-momentum expansion based on chiral symmetry 
can be constructed.                   

Naturally,                                            
the powers must be bounded from below for the expansion
to converge. This is in fact the case, 
with $\nu \geq 0$. 

Furthermore, the power formula 
Eq.~(\ref{eq_nu}) 
allows to predict
the leading orders of connected multi-nucleon forces.
Consider a $m$-nucleon irreducibly connected diagram
($m$-nucleon force) in an $A$-nucleon system ($m\leq A$).
The number of separately connected pieces is
$C=A-m+1$. Inserting this into
Eq.~(\ref{eq_nu}) together with $L=0$ and 
$\sum_i \Delta_i=0$ yields
$\nu=2m-4$. Thus, two-nucleon forces ($m=2$) appear
at $\nu=0$, three-nucleon forces ($m=3$) at
$\nu=2$ (but they happen to cancel at that order),
and four-nucleon forces at $\nu=4$ (they don't cancel).
More about this in the next sub-section.

For later purposes, we note that for an irreducible 
$NN$ diagram ($A=2$, $C=1$), the
power formula collapses to the very simple expression
\begin{equation}
\nu =  2L + \sum_i \Delta_i \,.
\label{eq_nunn}
\end{equation}

To summarize, at each order                            
$\nu$ we only have a well defined number of diagrams, 
which renders the theory feasible from a practical standpoint.
The magnitude of what has been left out at order $\nu$ can be estimated (in a 
very simple way) from 
$(Q/\Lambda_\chi)^{\nu+1}$. The ability to calculate observables (in 
principle) to any degree of accuracy gives the theory 
its predictive power.

\subsubsection{The ranking of nuclear forces}

\begin{figure}[t]\centering
%\vspace*{-0.5cm}
\scalebox{0.75}{\includegraphics{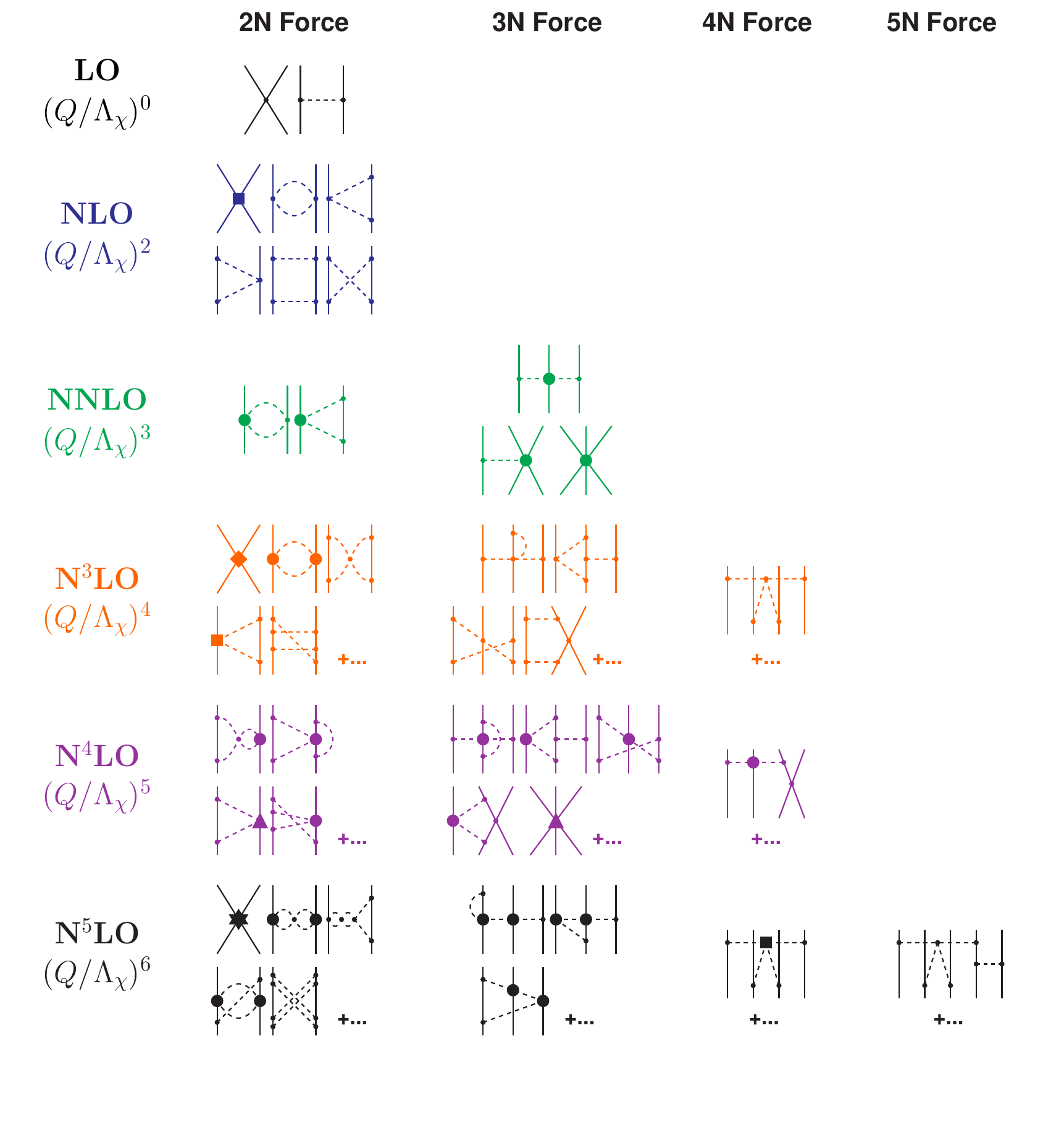}}
\vspace*{-1.5cm}
\caption{Hierarchy of nuclear forces in ChPT. Solid lines
represent nucleons and dashed lines pions. 
Small dots, large solid dots, solid squares, triangles, diamonds, and stars
denote vertices of index $\Delta_i = \, $ 0, 1, 2, 3, 4, and 6, respectively. 
Further explanations are
given in the text.}
\label{fig_hi}
\end{figure}

As shown in Fig.~\ref{fig_hi}, nuclear forces appear in ranked
orders in accordance with 
the power counting scheme. 

The lowest power is $\nu = 0$, also known as the leading order (LO).
At LO we have only two contact contributions with no momentum dependence 
($\sim Q^0$). They are signified by the 
four-nucleon-leg diagram 
with a small-dot vertex shown in the first row of 
Fig.~\ref{fig_hi}.
Besides this, we have the
static one-pion exchange (1PE), also shown 
in the first row of                 
Fig.~\ref{fig_hi}.              

In spite of its simplicity, this rough description                      
contains some of the main attributes of the $NN$ force. 
First, through the 1PE it generates the tensor component of the force
known to be crucial for the two-nucleon bound state. Second, 
it predicts correctly 
$NN$ phase parameters for high partial waves.               
At LO, the two terms which result from a partial-wave expansion of the contact term
impact states of zero orbital angular momentum and produce attraction at 
short- and intermediate-range.                              

Notice that there are no terms with power 
$\nu=1$, as they would violate parity conservation 
and time-reversal invariance.

The next order is then
$\nu=2$, next-to-leading order, or NLO.

Note that the two-pion exchange (2PE) makes its first appearance at this order,
and thus it is referred to as the 
``leading 2PE''. As is well known from decades of nuclear physics, 
this contribution is essential for a realistic account of the intermediate-range attraction.    
However, the leading 2PE has insufficient strength, for the following reason: 
the loops present in the diagrams which involve pions 
carry the power $\nu=2$ [cf.\ Eq.~(\ref{eq_nunn})],
and so only                                  
$\pi NN$ and $\pi \pi NN$ vertices with $\Delta_i = 0$ are allowed at this order. 
These vertices are known to be weak.
Moreover, seven new contacts appear at this order which 
impact $L = 0$ and $L = 1$ states. (As always, two-nucleon contact terms are indicated 
by four-nucleon-leg diagrams and a vertex of appropriate shape, in this case a solid square.) 
At this power, the appropriate operators                                  
include spin-orbit, central,
spin-spin, and tensor terms, namely all the spin and isospin
operator structures needed for a realistic description of the 
2NF, although the medium-range attraction still lacks 
sufficient strength.                             

At the next order, 
$\nu=3$ or next-to-next-to-leading order (NNLO), 
the 2PE contains the so-called          
$\pi\pi NN$ seagull vertices with two derivatives.                    
These vertices (proportional to 
the $c_i$ LECs,
Eq.~(\ref{eq_LD1}),
  and denoted by a large solid dot
in Fig.~\ref{fig_hi}), 
simulate correlated 2PE
and intermediate $\Delta(1232)$-isobar contributions.
Consistent with what the meson theory of                               
the nuclear forces~\cite{Lac80,MHE87} (cf.\ Sec.~\ref{sec_mesons}) has shown since a long time 
concerning the importance of these effects,            
at this order the 2PE finally provides medium-range     
attraction of realistic strength, bringing the description of the $NN$ force
to an almost quantitative level. 
No new contacts become available at NNLO. 

The discussion above reveals how                      
two- and many-nucleon forces are generated                      
and increase in number as we move to higher orders.
Three-nucleon forces appear at NLO,                          
but their net contribution vanishes at this order~\cite{Wei92}.
The first non-zero 3NF contribution is found 
at NNLO~\cite{Kol94,Epe02b}. It is therefore easy to understand why  
3NF are very weak as compared to the 2NF which contributes already at 
$(Q/\Lambda_\chi)^0$.

For $\nu =4$, or next-to-next-to-next-to-leading
order (N$^3$LO), we display some representative diagrams in 
Fig.~\ref{fig_hi}. There is a large attractive one-loop 2PE contribution (the bubble diagram with two large solid dots $\sim c_i^2$), which slightly over-estimates the 2NF attraction
at medium range. 
Two-pion-exchange graphs with two loops are seen at this order, together with 
three-pion exchange (3PE), which was determined to be very weak 
at N$^3$LO~\cite{Kai00a,Kai00b}.
The most important feature at this order is the presence 
of 15 additional contacts $\sim Q^4$, signified 
by the four-nucleon-leg diagram in the figure with the diamond-shaped vertex. 
These contacts impact states with orbital angular momentum up to $L = 2$, 
and are the reason for the                            
 quantitative description of the
two-nucleon force (up to approximately 300 MeV
in terms of laboratory energy) 
at this order~\cite{ME11,EM03}.
More 3NF diagrams show up 
at N$^3$LO, as well as the first contributions to 
four-nucleon forces (4NF).         
We then see that forces involving more and more nucleons appear for the
first time at higher and higher orders, which 
gives theoretical support to the fact that          
2NF $\gg$ 3NF $\gg$ 4NF
\ldots.

Further 2PE and 3PE occur at N$^4$LO (fifth order). The contribution to the 2NF 
at this order has been first calculated  by Entem {\it et al.}~\cite{Ent15a}. It turns out to be moderately repulsive, thus
compensating for the attractive surplus generated at N$^3$LO by the bubble diagram with two solid dots. The long- and intermediate-range 3NF contributions at this order have been evaluated~\cite{KGE12,KGE13}, but not yet applied in nuclear structure calculations. They are
expected to be sizeable. Moreover, a new set of 3NF contact terms appears~\cite{GKV11}.
The N$^4$LO 4NF has not been derived yet. Due to the subleading $\pi\pi N N$ seagull vertex (large solid dot $\sim c_i$), this 4NF could be sizeable.

Finally turning to N$^5$LO (sixth order): The dominant 2PE and 3PE contributions to the 2NF have been derived by Entem {\it et al.} in Ref.~\cite{Ent15b}, which represents
the most sophisticated investigation ever conducted in chiral EFT for the $NN$ system. The effects are small indicating the desired trend towards convergence of the chiral expansion for the 2NF. 
Moreover, a new set of 26 $NN$ contact terms $\sim Q^6$ occurs that contributes up to $F$-waves (represented by the $NN$ diagram with a star in Fig.~\ref{fig_hi})
bringing the total number of $NN$ contacts to 50~\cite{EM03a}.
The three-, four-, and five-nucleon forces of this order have not yet been derived.

\begin{figure}[t]\centering
\vspace*{-1.5cm}
%\hspace*{-1.7cm}
\scalebox{0.5}{\includegraphics{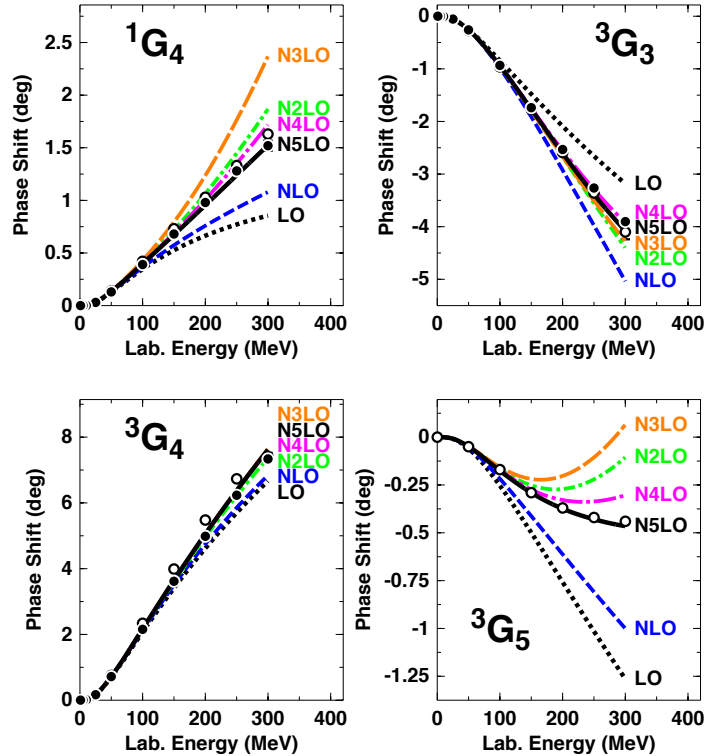}}
\vspace*{-2.0cm}
\caption{Phase-shifts of neutron-proton scattering in $G$ waves at all orders of ChPT
from LO to N$^5$LO.  
The filled and open circles represent the results from the Nijmegen multi-energy $np$ phase-shift analysis~\cite{Sto93} and the GWU single-energy $np$ analysis SP07~\cite{SP07}, respectively.}
\label{fig_ph6a}
\end{figure}

To summarize, we show in Fig.~\ref{fig_ph6a} the contributions to the phase shifts
of peripheral $NN$ scattering through 
 all orders from LO to N$^5$LO
 as obtained from a perturbative calculation.
   Note that the difference between the LO prediction 
(one-pion-exchange, dotted line) and the data (filled and open circles) is to 
be provided by two- and three-pion exchanges, i.e. the intermediate-range part
of the nuclear force. How well that is accomplished is a crucial test for
any theory of nuclear forces.  NLO produces only a small contribution, 
but N$^2$LO creates substantial intermediate-range attraction (most clearly 
seen in $^1G_4$, $^3G_5$). In fact, N$^2$LO is the largest 
contribution among all orders. This is due to the one-loop $2\pi$-exchange 
 triangle diagram which involves one $\pi\pi NN$-contact vertex 
proportional to $c_i$. As discussed,
the one-loop $2\pi$-exchange at N$^2$LO is attractive and 
 describes the intermediate-range attraction of the
nuclear force about right. At N$^3$LO, more one-loop 2PE is added by the 
bubble diagram with two $c_i$-vertices, a contribution that seemingly is 
overestimating the attraction. This attractive surplus is then compensated by
the prevailingly repulsive two-loop $2\pi$- and $3\pi$-exchanges that occur 
at N$^4$LO and N$^5$LO. 

In this context, it is worth noting that also in conventional meson 
theory~\cite{MHE87} (Sec.~\ref{sec_2pi}) the one-loop models for the 2PE contribution always show 
some excess of attraction (cf. Fig.~10 of Ref.~\cite{ME11}). The same is 
true for the dispersion theoretic approach pursued by the Paris 
group (see, e.~g., the predictions for $^1D_2$, $^3D_2$, and $^3D_3$
in Fig.~8 of Ref.~\cite{Vin79} which are all too attractive). 
In conventional meson theory, this
attraction is reduced by heavy-meson exchanges ($\rho$-, $\omega$-, and $\pi\rho$-exchange) 
which, however, has no place in chiral effective field theory (as a 
finite-range contribution). Instead, in the latter approach, two-loop 
$2\pi$- and $3\pi$-exchanges provide the corrective action.

\subsection{Quantitative chiral $NN$ potentials
\label{sec_pot}}

In the previous section, we mainly discussed the pion-exchange contributions to the $NN$
interaction.
They
describe the long- and intermediate-range
parts of the nuclear force, which are governed by chiral symmetry and rule the peripheral partial waves (cf.\ Fig.~\ref{fig_ph6a}).
However, for a ``complete'' nuclear force, we have to 
describe correctly all partial waves, including the lower ones.
In fact, in calculations of $NN$ observables at low energies (cross
sections, analyzing powers, etc.),
the partial
waves with $L\leq 2$ are the most important ones, generating
the largest contributions.
The same is true for microscopic nuclear structure
calculations.
The lower partial waves are dominated by the
dynamics at short distances.
Therefore, we need to look now more closely into the short-range part
of the $NN$ potential.

\subsubsection{$NN$ contact terms \label{sec_ct}}

In conventional meson theory (Sec.~\ref{sec_mesons}), the short-range nuclear force
is described by the exchange of heavy mesons, notably the
$\omega(782)$. 
Qualitatively, the short-distance behavior of the $NN$ 
potential is obtained
by Fourier transform of the propagator of
a heavy meson,
\begin{equation}
\int d^3q \frac{e^{i{\vec q} \cdot {\vec r}}}{m^2_\omega
+ {\vec q}^2} \;
\sim \;
 \frac{e^{-m_\omega r}}{r} \; .
\end{equation}

ChPT is an expansion in small momenta $Q$, too small
to resolve structures like a $\rho(770)$ or $\omega(782)$
meson, because $Q \ll \Lambda_\chi \approx m_{\rho,\omega}$.
But the latter relation allows us to expand the propagator 
of a heavy meson into a power series,
\begin{equation}
\frac{1}{m^2_\omega + Q^2} 
\approx 
\frac{1}{m^2_\omega} 
\left( 1 
- \frac{Q^2}{m^2_\omega}
+ \frac{Q^4}{m^4_\omega}
-+ \ldots
\right)
,
\label{eq_power}
\end{equation}
where the $\omega$ is representative
for any heavy meson of interest.
The above expansion suggests that it should be 
possible to describe the short distance part of
the nuclear force simply in terms of powers of
$Q/m_\omega$, which fits in well
with our over-all 
power expansion since $Q/m_\omega \approx Q/\Lambda_\chi$.
Since such terms act directly between nucleons, they are dubbed contact terms.

Contact terms play an important role in renormalization.
Regularization
of the loop integrals that occur in multi-pion exchange diagrams
typically generates polynomial terms
with coefficients that are, in part, infinite or scale
dependent (cf.\ Appendix B of Ref.~\cite{ME11}). Contact terms absorb infinities
and remove scale dependences, which is why they are also known as counter terms.

Due to parity, only even powers of $Q$
are allowed.
Thus, the expansion of the contact potential is
formally given by
\begin{equation}
V_{\rm ct} =
V_{\rm ct}^{(0)} + 
V_{\rm ct}^{(2)} + 
V_{\rm ct}^{(4)} + 
V_{\rm ct}^{(6)} 
+ \ldots \; ,
\label{eq_ct}
\end{equation}
where the supersript denotes the power or order.

We will now present, one by one, the various orders of 
$NN$ contact terms.\\

\paragraph{Zeroth order (LO)}
The contact Lagrangian
$\widehat{\cal L}^{(0)}_{NN}$,
Eq.~(\ref{eq_LNN0}), 
which is part of 
$\widehat{\cal L}^{\Delta=0}$,
Eq.~(\ref{eq_LD0}), 
leads to the following $NN$ contact potential,
\begin{equation}
V_{\rm ct}^{(0)}(\vec{p'},\vec{p}) =
C_S +
C_T \, \vec{\sigma}_1 \cdot \vec{\sigma}_2 \, ,
\label{eq_ct0}
\end{equation}
and, in terms of partial waves, we have
\be
V_{\rm ct}^{(0)}(^1 S_0)          &=&  \widetilde{C}_{^1 S_0} =
4\pi\, ( C_S - 3 \, C_T )
\nonumber \\
V_{\rm ct}^{(0)}(^3 S_1)          &=&  \widetilde{C}_{^3 S_1} =
4\pi\, ( C_S + C_T ) \,.
\label{eq_ct0_pw}
\ee
\\

\paragraph{Second order (NLO)}

The contact Lagrangian
$\widehat{\cal L}^{(2)}_{NN}$,
which is part of 
$\widehat{\cal L}^{\Delta=2}$,
Eq.~(\ref{eq_LD2}), 
generates the following $NN$ contact potential
\be
V_{\rm ct}^{(2)}(\vec{p'},\vec{p}) &=&
C_1 \, q^2 +
C_2 \, k^2 
\nonumber 
\\ &+& 
\left(
C_3 \, q^2 +
C_4 \, k^2 
\right) \vec{\sigma}_1 \cdot \vec{\sigma}_2 
\nonumber 
\\
&+& C_5 \left( -i \vec{S} \cdot (\vec{q} \times \vec{k}) \right)
\nonumber 
\\ &+& 
 C_6 \, ( \vec{\sigma}_1 \cdot \vec{q} )\,( \vec{\sigma}_2 \cdot 
\vec{q} )
\nonumber 
\\ &+& 
 C_7 \, ( \vec{\sigma}_1 \cdot \vec{k} )\,( \vec{\sigma}_2 \cdot 
\vec{k} ) \,,
\label{eq_ct2}
\ee

with partial-wave decomposition

\be
V_{\rm ct}^{(2)}(^1 S_0)          &=&  C_{^1 S_0} ( p^2 + {p'}^2 ) 
\nonumber \\
V_{\rm ct}^{(2)}(^3 P_0)          &=&  C_{^3 P_0} \, p p'
\nonumber \\
V_{\rm ct}^{(2)}(^1 P_1)          &=&  C_{^1 P_1} \, p p' 
\nonumber \\
V_{\rm ct}^{(2)}(^3 P_1)          &=&  C_{^3 P_1} \, p p' 
\nonumber \\
V_{\rm ct}^{(2)}(^3 S_1)          &=&  C_{^3 S_1} ( p^2 + {p'}^2 ) 
\nonumber \\
V_{\rm ct}^{(2)}(^3 S_1- ^3 D_1)  &=&  C_{^3 S_1- ^3 D_1}  p^2 
\nonumber \\
V_{\rm ct}^{(2)}(^3 D_1- ^3 S_1)  &=&  C_{^3 S_1- ^3 D_1}  {p'}^2 
\nonumber \\
V_{\rm ct}^{(2)}(^3 P_2)          &=&  C_{^3 P_2} \, p p' 
\label{eq_ct2_pw}
\ee
which obviously contributes up to $P$ waves.
\\

\paragraph{Fourth order (N$^3$LO)}

The contact potential of order four reads
\be
V_{\rm ct}^{(4)}(\vec{p'},\vec{p}) &=&
D_1 \, q^4 +
D_2 \, k^4 +
D_3 \, q^2 k^2 +
D_4 \, (\vec{q} \times \vec{k})^2 
\nonumber 
\\ &+& 
\left(
D_5 \, q^4 +
D_6 \, k^4 +
D_7 \, q^2 k^2 +
D_8 \, (\vec{q} \times \vec{k})^2 
\right) \vec{\sigma}_1 \cdot \vec{\sigma}_2 
\nonumber 
\\ &+& 
\left(
D_9 \, q^2 +
D_{10} \, k^2 
\right) \left( -i \vec{S} \cdot (\vec{q} \times \vec{k}) \right)
\nonumber 
\\ &+& 
\left(
D_{11} \, q^2 +
D_{12} \, k^2 
\right) ( \vec{\sigma}_1 \cdot \vec{q} )\,( \vec{\sigma}_2 
\cdot \vec{q})
\nonumber 
\\ &+& 
\left(
D_{13} \, q^2 +
D_{14} \, k^2 
\right) ( \vec{\sigma}_1 \cdot \vec{k} )\,( \vec{\sigma}_2 
\cdot \vec{k})
\nonumber 
\\ &+& 
D_{15} \left( 
\vec{\sigma}_1 \cdot (\vec{q} \times \vec{k}) \, \,
\vec{\sigma}_2 \cdot (\vec{q} \times \vec{k}) 
\right) .
\label{eq_ct4}
\ee
The rather lengthy partial-wave expressions of this order
are given in Appendix E of Ref.~\cite{ME11}. These contacts affect partial waves up to $D$ waves.\\

\paragraph{Sixth order (N$^5$LO)}

At sixth order, 26 new contact terms appear, bringing the total number to 50. These terms
as well as their partial-wave decomposition have been worked out in Ref.~\cite{EM03a}.
They contribute up to $F$-waves.
So far, these terms have not been used in the construction of $NN$ potentials.

\subsubsection{Definition of $NN$ potential \label{sec_pot1}}

We have now rounded up everything needed for a realistic
nuclear force---long, intermediate, and short ranged 
components---and so we can finally proceed to the lower
partial waves. However, here we encounter another problem.
The two-nucleon system at low angular momentum, particularly
in $S$ waves, is characterized by the
presence of a shallow bound state (the deuteron)
and large scattering lengths.
Thus, perturbation theory does not apply.
In contrast to $\pi$-$\pi$ and $\pi$-$N$,
the interaction between nucleons is not suppressed
in the chiral limit ($Q\rightarrow 0$).
Weinberg~\cite{Wei91} showed that the strong enhancement of the
scattering amplitude arises from purely nucleonic intermediate
states (``infrared enhancement''). He therefore suggested to use perturbation theory to
calculate the $NN$ potential (i.e., the irreducible graphs) and to apply this potential
in a scattering equation 
to obtain the $NN$ amplitude. We will follow
this prescription.

The potential $V$ as discussed in previous sections is, in principal, an invariant amplitude and, thus, satisfies a relativistic scattering equation, for which we choose the
BbS equation~\cite{BS66}, which reads explicitly,
\begin{equation}
{T}({\vec p}~',{\vec p})= {V}({\vec p}~',{\vec p})+
\int \frac{d^3p''}{(2\pi)^3} \:
{V}({\vec p}~',{\vec p}~'') \:
\frac{M_N^2}{E_{p''}} \:  
\frac{1}
{{ p}^{2}-{p''}^{2}+i\epsilon} \:
{T}({\vec p}~'',{\vec p}) 
\label{eq_bbs2}
\end{equation}
with $E_{p''}\equiv \sqrt{M_N^2 + {p''}^2}$.
The advantage of using a relativistic scattering equation is that it automatically
includes relativistic corrections to all orders. Thus, in the scattering equation,
no propagator modifications are necessary when raising the order to which the
calculation is conducted.

Defining
\begin{equation}
\widehat{V}({\vec p}~',{\vec p})
\equiv 
\frac{1}{(2\pi)^3}
\sqrt{\frac{M_N}{E_{p'}}}\:  
{V}({\vec p}~',{\vec p})\:
 \sqrt{\frac{M_N}{E_{p}}}
\label{eq_minrel1}
\end{equation}
and
\begin{equation}
\widehat{T}({\vec p}~',{\vec p})
\equiv 
\frac{1}{(2\pi)^3}
\sqrt{\frac{M_N}{E_{p'}}}\:  
{T}({\vec p}~',{\vec p})\:
 \sqrt{\frac{M_N}{E_{p}}}
\,,
\label{eq_minrel2}
\end{equation}
where the factor $1/(2\pi)^3$ is added for convenience,
the BbS equation collapses into the usual, nonrelativistic
Lippmann-Schwinger (LS) equation,
\begin{equation}
 \widehat{T}({\vec p}~',{\vec p})= \widehat{V}({\vec p}~',{\vec p})+
\int d^3p''\:
\widehat{V}({\vec p}~',{\vec p}~'')\:
\frac{M_N}
{{ p}^{2}-{p''}^{2}+i\epsilon}\:
\widehat{T}({\vec p}~'',{\vec p}) \, .
\label{eq_LS}
\end{equation}
Since 
$\widehat V$ 
satisfies Eq.~(\ref{eq_LS}), 
it can be used like a nonrelativistic potential, and 
$\widehat{T}$ 
may be perceived as the conventional nonrelativistic 
$T$-matrix.

\subsubsection{Regularization and non-perturbative renormalization}
\label{sec_reno}

Iteration of $\widehat V$ in the LS equation, Eq.~(\ref{eq_LS}),
requires cutting $\widehat V$ off for high momenta to avoid infinities.
This is consistent with the fact that ChPT
is a low-momentum expansion which
is valid only for momenta $Q \ll \Lambda_\chi \approx 1$ GeV.
Therefore, the potential $\widehat V$
is multiplied
with the regulator function $f(p',p)$,
\begin{equation}
{\widehat V}(\vec{ p}~',{\vec p})
\longmapsto
{\widehat V}(\vec{ p}~',{\vec p}) \, f(p',p) 
\end{equation}
with
\begin{equation}
f(p',p) = \exp[-(p'/\Lambda)^{2n}-(p/\Lambda)^{2n}] \,,
\label{eq_f}
\end{equation}
such that
\begin{equation}
{\widehat V}(\vec{ p}~',{\vec p}) \, f(p',p) 
\approx
{\widehat V}(\vec{ p}~',{\vec p})
\left\{1-\left[\left(\frac{p'}{\Lambda}\right)^{2n}
+\left(\frac{p}{\Lambda}\right)^{2n}\right]+ \ldots \right\} 
\,.
\label{eq_reg_exp}
\end{equation}
Typical choices for the cutoff parameter $\Lambda$ that
appears in the regulator are 
$\Lambda \approx 0.5 \mbox{ GeV} < \Lambda_\chi \approx 1$ GeV.

Equation~(\ref{eq_reg_exp}) provides an indication of the fact that
the exponential cutoff does not necessarily
affect the given order at which 
the calculation is conducted.
For sufficiently large $n$, the regulator introduces contributions that 
are beyond the given order. Assuming a good rate
of convergence of the chiral expansion, such orders are small 
as compared to the given order and, thus, do not
affect the accuracy at the given order.
In calculations, one uses, of course,
the exponential form, Eq.~(\ref{eq_f}),
and not the expansion Eq.~(\ref{eq_reg_exp}).

It is pretty obvious that results for the $T$-matrix may
depend sensitively on the regulator and its cutoff parameter.
This is acceptable if one wishes to build models.
For example, the meson models of the past~\cite{Mac89,MHE87} (Sec.~\ref{sec_mesons})
always depended sensitively on the choices for the
cutoff parameters, and they were
welcome as additional fit parameters to further improve the reproduction of the $NN$ data.
However, the EFT approach wishes to be more fundamental
in nature and not just another model.

In field theories, divergent integrals are not uncommon and methods have
been devised for how to deal with them.
One regulates the integrals and then removes the dependence
on the regularization parameters (scales, cutoffs)
by renormalization. In the end, the theory and its
predictions do not depend on cutoffs
or renormalization scales.

Renormalizable quantum field theories, like QED,
have essentially one set of prescriptions 
that takes care of renormalization through all orders. 
In contrast, 
EFTs are renormalized order by order, i.~e., each order comes with the counter terms
needed to renormalize that order. 
Note that this applies only to perturbative calculations. 
The $NN$ {\it potential} is calculated perturbatively and hence properly renormalized.

 However, the story is different for the $NN$ {\it amplitude} ($T$-matrix) that results from a solution of the  LS equation, Eq.~(\ref{eq_LS}),  which is a {\it nonperturbative} resummation of the potential.
This resummation is necessary in {\it nuclear} EFT because
nuclear physics is characterized by bound states which
are nonperturbative in nature.
EFT power counting may be different for nonperturbative processes as
compared to perturbative ones. Such difference may be caused by the infrared
enhancement of the reducible diagrams generated in the LS equation.

Weinberg's implicit assumption~\cite{Wei90,Wei09} was that the counterterms
introduced to renormalize the perturbatively calculated
potential, based upon naive dimensional analysis (``Weinberg counting''),
are also sufficient to renormalize the nonperturbative
resummation of the potential in the LS equation.
In 1996, Kaplan, Savage, and Wise (KSW)~\cite{KSW96}
pointed out that there are problems with the Weinberg scheme
if the LS equation is renormalized 
by minimally-subtracted dimensional regularization.
This criticism resulted in a flurry of publications on
the renormalization of the nonperturbative
$NN$ problem and I like to refer the interested reader to Ref.~\cite{ME11}
for a comprehensive consideration of the issue.

\subsubsection{$NN$ potentials order by order}
\label{sec_pot2}

\begin{figure}[t]\centering

%\vspace*{-0.5cm}
%\hspace*{-0.4cm}
%\scalebox{0.35}{\includegraphics{figph1a.pdf}}
%\hspace*{-0.5cm}
%\scalebox{0.35}{\includegraphics{figph1b.pdf}}
%\vspace*{-1.0cm}

\vspace*{-1cm}
\scalebox{0.5}{\includegraphics{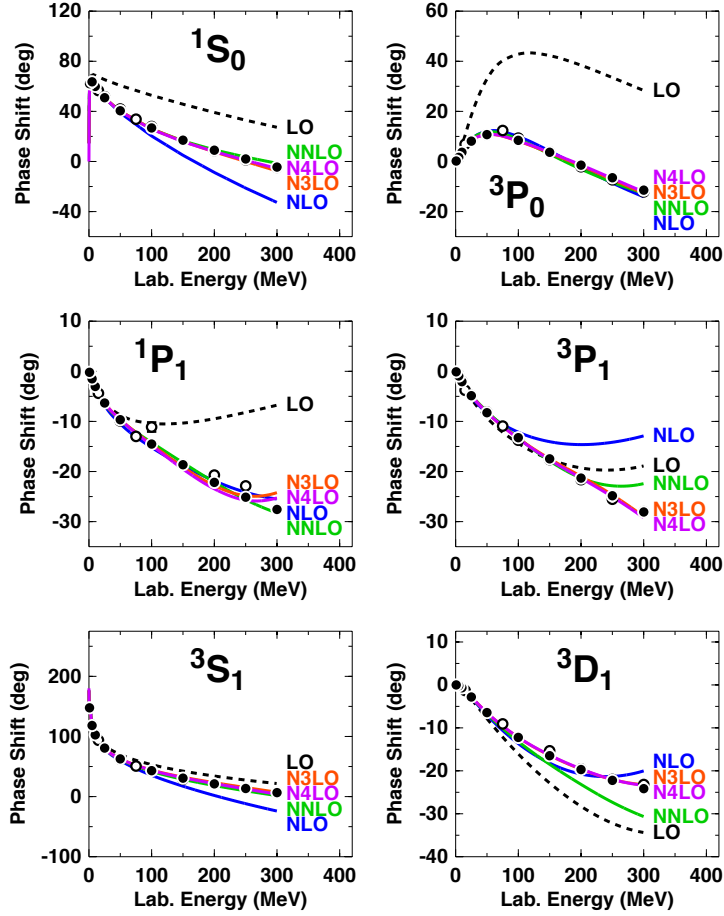}}
\vspace*{-1.0cm}
\caption{
Chiral expansion of neutron-proton scattering as represented by the phase shifts 
for $J \leq 1$.
 Five orders ranging from LO to N$^4$LO are shown as 
denoted.
Filled and open circles as in Fig.~\ref{fig_ph6a}.
(From Ref.~\cite{EMN17}.)
\label{fig_ph1a}}
\end{figure}

\begin{figure}[t]\centering
\vspace*{-1.0cm}
\scalebox{0.5}{\includegraphics{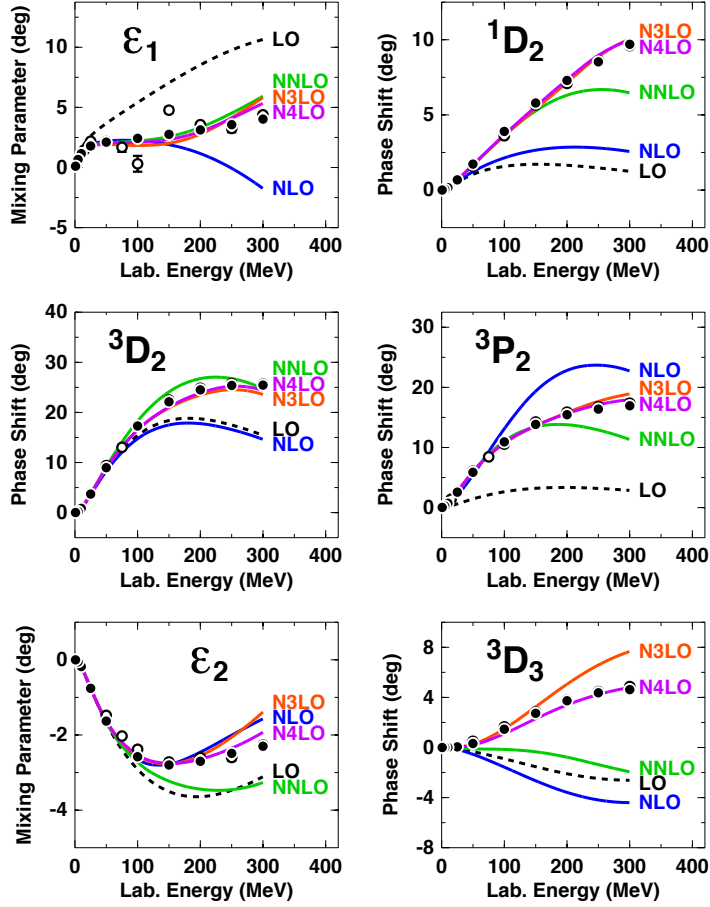}}
\vspace*{-1.0cm}
\caption{Same as Fig.~\ref{fig_ph1a}, but for $^3P_2$, $D$-waves, and mixing parameters 
$\epsilon_1$ and $\epsilon_2$.
\label{fig_ph1b}}
\end{figure}

\begin{table}[t]
\caption{$\chi^2/$datum for the fit of the 2016 $NN$ data base by $NN$ potentials at various orders of chiral EFT. (From Ref.~\cite{EMN17})
\label{tab_chi}}
\smallskip
\begin{tabular*}{\textwidth}{@{\extracolsep{\fill}}ccccccc}
\hline 
\hline 
\noalign{\smallskip}
 $T_{\rm lab}$ bin (MeV) & No.\ of data & LO & NLO & NNLO & N$^3$LO & N$^4$LO \\
\hline
\noalign{\smallskip}
\multicolumn{7}{c}{\bf proton-proton} \\
0--100 & 795 & 520 & 18.9  & 2.28   &  1.18 & 1.09 \\
0--190 & 1206  & 430 & 43.6  &  4.64 & 1.69 & 1.12 \\
0--290 & 2132 & 360 & 70.8  & 7.60  &  2.09  & 1.21 \\
\hline
\noalign{\smallskip}
\multicolumn{7}{c}{\bf neutron-proton} \\
0--100 & 1180 & 114 & 7.2  &  1.38  & 0.93  & 0.94 \\
0--190 & 1697 &  96 & 23.1  &  2.29 &  1.10  & 1.06 \\
0--290 & 2721 &  94 &  36.7 & 5.28  &  1.27 & 1.10 \\
\hline
\noalign{\smallskip}
\multicolumn{7}{c}{\boldmath $pp$ plus $np$} \\
0--100 & 1975 & 283 &  11.9 &  1.74  & 1.03   & 1.00  \\
0--190 & 2903 & 235 &  31.6 &  3.27 & 1.35   &  1.08 \\
0--290 & 4853 & 206 &  51.5 & 6.30  & 1.63   &  1.15 \\
\hline
\hline
\noalign{\smallskip}
\end{tabular*}
\end{table}

$NN$ potentials depend on two different sets of parameters, the $\pi N$ and the $NN$ LECs.
The $\pi N$ LECs are the coefficients that appear in the $\pi N$ Langrangians, e.~g., the 
$c_i$ in Eq.~(\ref{eq_LD1}). They are determined in $\pi N$ analysis~\cite{Hof15}.
The $NN$ LECs are the coefficients of the $NN$ contact terms (cf.\ Sec.~\ref{sec_ct}).
They are fixed by an optimal fit to the $NN$ data below pion-production threshold, see
Ref.~\cite{EMN17} for details.

$NN$ potentials are then constructed order by order
and the accuracy improves as the order increases.
How well the chiral expansion converges in 
important lower partial waves
is demonstrated in Figs.~\ref{fig_ph1a} and \ref{fig_ph1b},
where we show phase parameters for potentials developed
through all orders from LO to N$^4$LO~\cite{EMN17}.~\footnote{Other chiral $NN$ potentials can be found in Refs.~\cite{ORK96,EGM05,EKM15a,EKM15b,Eks13,Gez14,Pia15,Pia16,PAA15}.}
These figures clearly reveal
substantial improvements in the reproduction of the empirical
phase shifts with increasing order.

The $\chi^2$/datum for the reproduction of the $NN$ data at various orders of chiral EFT 
are shown in Table~\ref{tab_chi} for different energy intervals below 290 MeV laboratory energy ($T_{\rm lab}$). 
The bottom line of Table~\ref{tab_chi} summarizes the essential results.
For the close to 5000 $pp$ plus $np$ data below 290 MeV (pion-production threshold),
the $\chi^2$/datum 
is 51.4 at NLO and 6.3 at NNLO. Note that the number of $NN$ contact terms is the same for both orders. The improvement is entirely due to an improved description of the 2PE contribution, which is responsible for the crucial intermediate-range attraction of the nuclear force.
At NLO, only the uncorrelated 2PE is taken into account which is insufficient. From the classic meson-theory of nuclear forces~\cite{MHE87} (Sec.~\ref{sec_mesons}), it is wellknown that $\pi$-$\pi$ correlations and nucleon resonances need to be taken into account for a realistic model of 2PE.
As discussed, in the chiral theory, these contributions are encoded in the subleading $\pi N$ vertexes with LECs denoted by $c_i$, Eq.~(\ref{eq_LD1}).
These enter at NNLO and are the reason for the substantial improvements we encounter at that order.

To continue on the bottom line of Table~\ref{tab_chi}, after NNLO,
the $\chi^2$/datum then further improves to 1.63 at N$^3$LO and, finally, reaches the almost perfect value of 1.15 at N$^4$LO---a fantastic convergence.

\begin{table}[t]
\small
\caption{Two- and three-nucleon bound-state properties as predicted by
  $NN$ potentials at various orders of chiral EFT ($\Lambda = 500$ MeV in all cases).
(Deuteron: Binding energy $B_d$, asymptotic $S$ state $A_S$,
asymptotic $D/S$ state $\eta$, structure radius $r_{\rm str}$,
quadrupole moment $Q$, $D$-state probability $P_D$; the predicted
$r_{\rm str}$ and $Q$ are without meson-exchange current contributions
and relativistic corrections. Triton: Binding energy $B_t$.)
$B_d$ is fitted, all other quantities are predictions. (From Ref.~\cite{EMN17}.)
\label{tab_deu}}
\smallskip
\begin{tabular*}{\textwidth}{@{\extracolsep{\fill}}lllllll}
\hline 
\hline 
\noalign{\smallskip}
 & LO & NLO & NNLO & N$^3$LO & N$^4$LO & Empirical$^a$ \\
\hline
\noalign{\smallskip}
{\bf Deuteron} \\
$B_d$ (MeV) &
 2.224575& 2.224575 &
 2.224575 & 2.224575 & 2.224575 & 2.224575(9) \\
$A_S$ (fm$^{-1/2}$) &
 0.8526& 0.8828 &
0.8844 & 0.8853 & 0.8852 & 0.8846(9)  \\
$\eta$         & 
 0.0302& 0.0262 &
0.0257& 0.0257 & 0.0258 & 0.0256(4) \\
$r_{\rm str}$ (fm)   & 1.911
      & 1.971 & 1.968
       & 1.970
       & 1.973 &
 1.97507(78) \\
$Q$ (fm$^2$) &
 0.310& 0.273&
 0.273 & 
 0.271 & 0.273 &
 0.2859(3)  \\
$P_D$ (\%)    & 
 7.29& 3.40&
4.49 & 4.15 & 4.10 & --- \\
\hline
\noalign{\smallskip}
{\bf Triton} \\
$B_t$ (MeV) & 11.09  & 8.31  & 8.21 & 8.09  & 8.08 & 8.48 \\
\hline
\hline
\noalign{\smallskip}
\end{tabular*}
\footnotesize
$^a$See Table XVIII of Ref.~\cite{Mac01} for references;
the empirical value for $r_{\rm str}$ is from Ref.~\cite{Jen11}.\\
\end{table}

The evolution of the deuteron properties from LO to N$^4$LO  of chiral EFT are shown in Table~\ref{tab_deu}.
In all cases, we fit the deuteron binding energy to its empirical value of 2.224575 MeV
using the non-derivative $^3S_1$ contact. All other deuteron properties are predictions.
Already at NNLO, the deuteron has converged to its empirical properties and stays there
through the higher orders.

At the bottom of Table~\ref{tab_deu}, we also show the predictions for the triton binding
as obtained in 34-channel charge-dependent Faddeev calculations using only 2NFs. The results show smooth and steady convergence, order by order, towards a value around 8.1 MeV that is reached at the highest orders shown. This contribution from the 2NF will require only a moderate 3NF. 
The relatively low deuteron $D$-state probabilities ($\approx 4.1$\% at N$^3$LO and N$^4$LO) and the concomitant generous triton binding energy predictions are
a reflection of the fact that our $NN$ potentials are soft (which is, at least in part, due to their non-local character). 

\subsection{Nuclear many-body forces \label{sec_manyNF}}

Two-nucleon forces derived from chiral EFT 
have been applied, often successfully, in the many-body system.                                  
On the other hand, over the past several years we have learnt that, for some few-nucleon
reactions and nuclear structure issues, 3NFs cannot be neglected.              
The most well-known cases are the so-called $A_y$ puzzle of $N$-$d$ scattering~\cite{EMW02},
the ground state of $^{10}$B~\cite{Cau02}, and the saturation of nuclear matter~\cite{Cor14,Sam15,MS16}.
As we observed previously, 
the EFT approach generates          
consistent two- and many-nucleon forces in a natural way 
(cf.\ the overview given in Fig.~\ref{fig_hi}).
We now shift our focus to chiral three- and four-nucleon forces.

\subsubsection{Three-nucleon forces}
\label{sec_3nfs}

Weinberg~\cite{Wei92} was the first to discuss     
nuclear three-body forces in the context of ChPT. Not long after that, 
the first 3NF at NNLO was derived by van Kolck~\cite{Kol94}.

For a 3NF, we have $A=3$ and $C=1$ and, thus, Eq.~(\ref{eq_nu})
implies
\begin{equation}
\nu = 2 + 2L + 
\sum_i \Delta_i \,.
\label{eq_nu3nf}
\end{equation}
We will use this equation to analyze 3NF contributions
order by order.

\paragraph{Next-to-leading order}

The lowest possible power is obviously $\nu=2$ (NLO), which
is obtained for no loops ($L=0$) and 
only leading vertices
($\sum_i \Delta_i = 0$). 
As discussed by Weinberg~\cite{Wei92} and van Kolck~\cite{Kol94}, 
the contributions from these diagrams
vanish at NLO. So, the bottom line is that there is no genuine 3NF contribution at NLO.
The first non-vanishing 3NF appears at NNLO.

\paragraph{Next-to-next-to-leading order}

\begin{figure}[t]\centering
%\vspace{-10.0cm}
%\hspace*{-2.0cm}
\scalebox{0.7}{\includegraphics{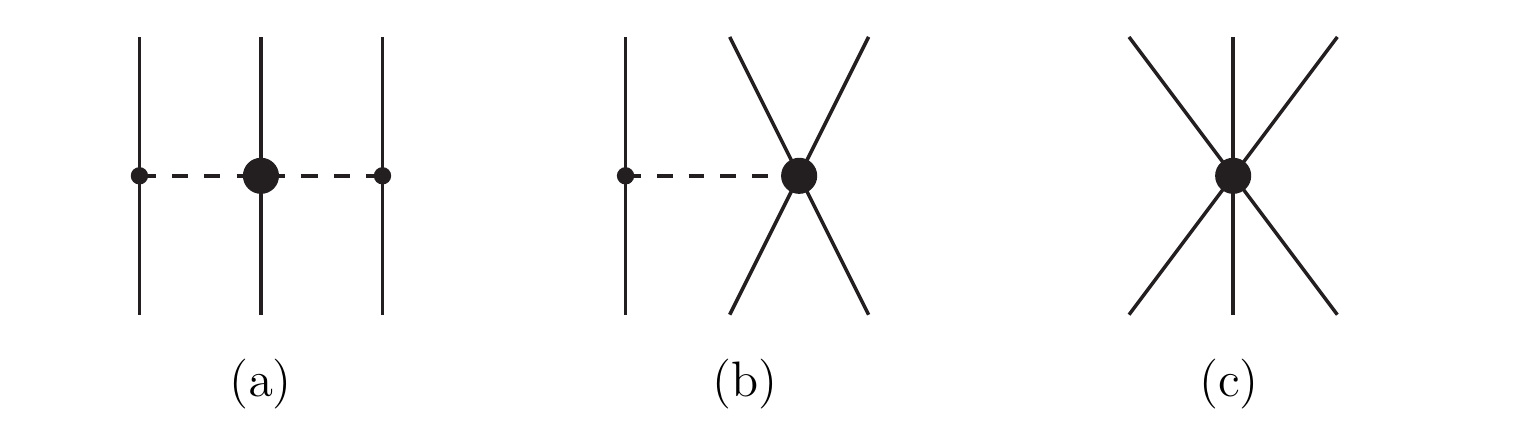}}
\vspace{-0.5cm}
\caption{The three-nucleon force at NNLO
with (a) 2PE, (b) 1PE, and (c) contact diagrams.
Notation as in Fig.~\ref{fig_hi}.}
\label{fig_3nf_nnlo}
\end{figure}

The power $\nu=3$ (NNLO) is obtained when
there are no loops ($L=0$) and 
$\sum_i \Delta_i = 1$, i.e., 
$\Delta_i=1$ for one vertex 
while $\Delta_i=0$ for all other vertices.
There are three topologies which fulfill this condition,
known as the 2PE, 1PE,
and contact graphs~\cite{Kol94,Epe02b}
(Fig.~\ref{fig_3nf_nnlo}).

The 2PE 3N-potential is derived to be
\begin{equation}
V^{\rm 3NF}_{\rm 2PE} = 
\left( \frac{g_A}{2f_\pi} \right)^2
\frac12 
\sum_{i \neq j \neq k}
\frac{
( \vec \sigma_i \cdot \vec q_i ) 
( \vec \sigma_j \cdot \vec q_j ) }{
( q^2_i + m^2_\pi )
( q^2_j + m^2_\pi ) } \;
F^{ab}_{ijk} \;
\tau^a_i \tau^b_j
\label{eq_3nf_nnloa}
\end{equation}
with $\vec q_i \equiv \vec{p_i}' - \vec p_i$, 
where 
$\vec p_i$ and $\vec{p_i}'$ are the initial
and final momenta of nucleon $i$, respectively, and
\begin{equation}
F^{ab}_{ijk} = \delta^{ab}
\left[ - \frac{4c_1 m^2_\pi}{f^2_\pi}
+ \frac{2c_3}{f^2_\pi} \; \vec q_i \cdot \vec q_j \right]
+ 
\frac{c_4}{f^2_\pi}  
\sum_{c} 
\epsilon^{abc} \;
\tau^c_k \; \vec \sigma_k \cdot [ \vec q_i \times \vec q_j] \; .
\label{eq_3nf_nnlob}
\end{equation}  

It is interesting to observe that there are clear analogies between this force and earlier          
2PE 3NFs already proposed decades ago, particularly the Fujita-Miyazawa~\cite{FM57} and
the Tucson-Melbourne (TM)~\cite{Coo79} forces. 

The 2PE 3NF does not introduce additional fitting constants, 
since the LECs $c_1$, $c_3$, and $c_4$ are already present in the 2PE 2NF.

The other two 3NF contributions shown in Fig.~\ref{fig_3nf_nnlo}
are easily derived by taking the last two terms of the $\Delta=1$ Langrangian, Eq.~(\ref{eq_LD1}),
into account. The 1PE contribution is
\begin{equation}
V^{\rm 3NF}_{\rm 1PE} = 
-D \; \frac{g_A}{8f^2_\pi} 
\sum_{i \neq j \neq k}
\frac{\vec \sigma_j \cdot \vec q_j}{
 q^2_j + m^2_\pi }
( \mbox{\boldmath $\tau$}_i \cdot \mbox{\boldmath $\tau$}_j ) 
( \vec \sigma_i \cdot \vec q_j ) 
\label{eq_3nf_nnloc}
\end{equation}
and the 3N contact potential is given by 
\begin{equation}
V^{\rm 3NF}_{\rm ct} = E \; \frac12
\sum_{i \neq j \neq k}
 \mbox{\boldmath $\tau$}_i \cdot \mbox{\boldmath $\tau$}_j  \; .
\label{eq_3nf_nnlod}
\end{equation}
These 3NF potentials introduce 
two additional constants, $D$ and $E$, which can be constrained in             
 more than one way.                                    
One may use 
the triton binding energy and the $nd$ doublet scattering
length $^2a_{nd}$~\cite{Epe02b}
or an optimal global fit of the properties of light nuclei~\cite{Nav07}.
 Alternative choices include 
the binding energies of $^3$H and $^4$He~\cite{Nog06} or
the binding energy of $^3$H and the point charge radius of $^4$He~\cite{Heb11}.
Another method makes use of
the triton binding energy and the Gamow-Teller matrix element of tritium $\beta$-decay~\cite{Mar12}.
When the values of $D$ and $E$ are determined, the results for other
observables involving three or more nucleons are true theoretical predictions.

Applications of the leading 3NF include few-nucleon 
reactions~\cite{Epe02b,NRQ10,Viv13}, structure of light- and medium-mass nuclei~\cite{Hag12a,Hag12b,BNV13,Her13,Hag14a,Sim17,Mor17}, and infinite matter~\cite{HS10,Heb11,Hag14b,Cor13,Cor14,Sam15,MS16},
often with 
satisfactory results. Some problems, though, remain unresolved, such as 
the well-known `$A_y$ puzzle' in nucleon-deuteron                
scattering~\cite{EMW02,Epe02b}.
Predictions which employ only 2NFs underestimate 
the analyzing power in $p$-$^3$He scattering
to a larger degree than in $p$-$d$. 
Although the $p$-$^3$He $A_y$ improves considerably (more than in the $p$-$d$ case) when the leading 3NF is          
included~\cite{Viv13}, the disagreement with the data is not fully removed. 
Also, predictions for light nuclei are not quite satisfactory.

In summary, the leading 3NF of ChPT is a remarkable contribution. It gives validation to, 
and provides a better framework for, 
3NFs which were proposed already five decades ago; it alleviates existing problems            
in few-nucleon reactions and the spectra of
light nuclei.
Nevertheless, we still face several challenges.                   
With regard to the 2NF, we have discussed earlier that it is necessary 
to go to order four or even five for convergence and high-precison predictions. 
Thus, the 3NF at N$^3$LO must be considered simply as a matter of consistency with the 2NF sector. 
At the same time, one hopes that its 
inclusion may result in further improvements with the aforementioned unresolved problems.

\begin{figure}[t]\centering
\vspace*{-0.5cm}
%\hspace*{-1.5cm}
\scalebox{0.9}{\includegraphics{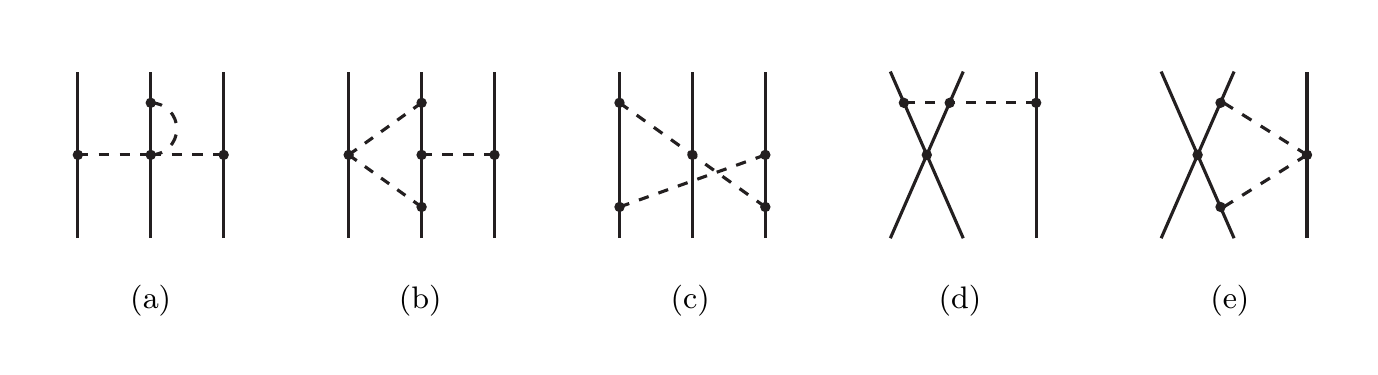}}
\vspace*{-0.75cm}
\caption{Leading one-loop 3NF diagrams at N$^3$LO.
We show one representative example for each of five topologies,
which are: (a) 2PE, (b) 1PE-2PE, (c) ring, (d) contact-1PE, (e) contact-2PE.
Notation as in Fig.~\ref{fig_hi}.}
\label{fig_3nf_n3lo}
\end{figure}

\paragraph{Next-to-next-to-next-to-leading order}
\label{sec_3nfn3lo}

At N$^3$LO, there are loop and tree diagrams.
For the loops (Fig.~\ref{fig_3nf_n3lo}), we have
$L=1$ and, therefore, all $\Delta_i$ have to be zero
to ensure $\nu=4$. 
Thus, these one-loop 3NF diagrams can include
only leading order vertices, the parameters of which
are fixed from $\pi N$ and $NN$ analysis.
The diagrams
have been evaluated by the Bochum-Bonn group~\cite{Ber08,Ber11}.
The long-range part of the chiral N$^3$LO 3NF has been
tested in the triton and in three-nucleon scattering~\cite{Gol14}
yielding only moderate effects. The long- and short-range parts of this
force have been used in neutron matter calculations
(together with the N$^3$LO 4NF) producing relatively large contributions
from the 3NF~\cite{Kru13,Dri16}. Thus, the ultimate assessment of the N$^3$LO 3NF is still
outstanding and will require more few- and many-body applications.

\paragraph{The 3NF at N$^4$LO}
\label{sec_3nfn4lo}

\begin{figure}[t]\centering
\vspace*{-0.5cm}
\scalebox{0.9}{\includegraphics{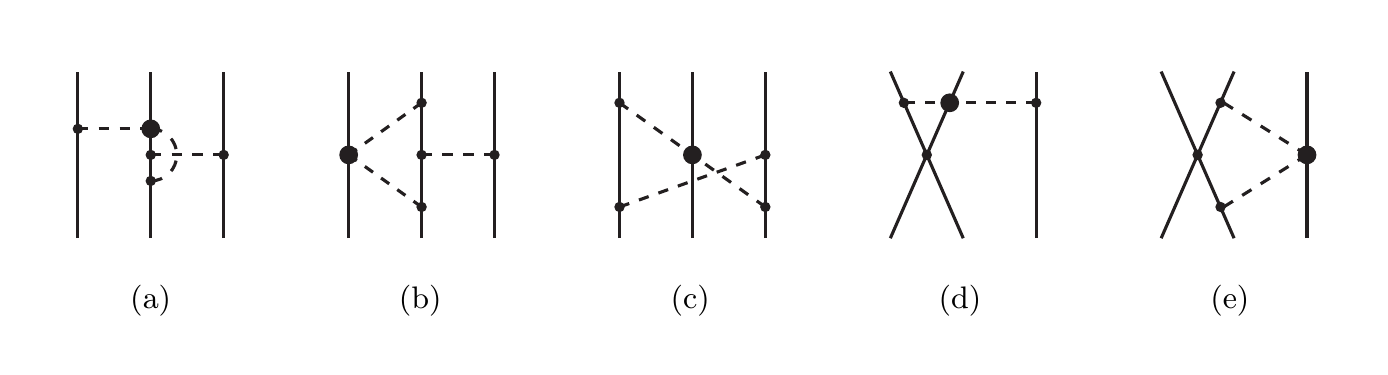}}
\vspace*{-0.75cm}
\caption{Sub-leading one-loop 3NF diagrams which appear at N$^4$LO
with topologies similar to Fig.~\ref{fig_3nf_n3lo}.
Notation as in Fig.~\ref{fig_hi}.}
\label{fig_3nf_n4loloops}
\end{figure}

\begin{figure}[t]\centering
\vspace*{-0.5cm}
\scalebox{1.0}{\includegraphics{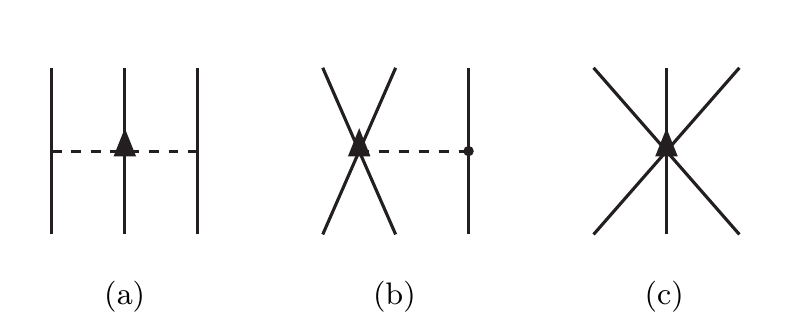}}
\vspace*{-0.3cm}
\caption{3NF tree graphs at N$^4$LO ($\nu=5$) denoted by: (a) 2PE, (b) 1PE-contact, and (c) contact. Notation as in Fig.~\ref{fig_hi}.}
\label{fig_3nf_n4lotrees}
\end{figure}

In the meantime, one may go ahead and look
at the next order of 3NFs, which is N$^4$LO or $\nu=5$.
The loop contributions that occur at this order
are obtained by replacing in the N$^3$LO loops
one vertex by a $\Delta_i=1$ vertex (with LEC $c_i$), Fig.~\ref{fig_3nf_n4loloops},
which is why these loops may be more sizable than the N$^3$LO loops.
The 2PE, 1PE-2PE, and ring topologies have been evaluated~\cite{KGE12,KGE13} so far.
In addition, we have three `tree' topologies (Fig.~\ref{fig_3nf_n4lotrees}), which include
a new set of 3N contact interactions that has recently been derived
by the Pisa group~\cite{GKV11}. Contact terms are typically simple (as compared
to loop diagrams) and their coefficients are essentially free. 
Therefore, it is an
attractive project to test some terms (in particular, the spin-orbit terms) 
of the N$^4$LO contact 3NF in calculations of few-body reactions (specifically,
the $p$-$d$ and $p$-$^3$He $A_y$), which is under way~\cite{Gir16}.

\subsubsection{Four-nucleon forces}

\begin{figure}[t]\centering
%\vspace*{-1.5cm}
\scalebox{0.78}{\includegraphics{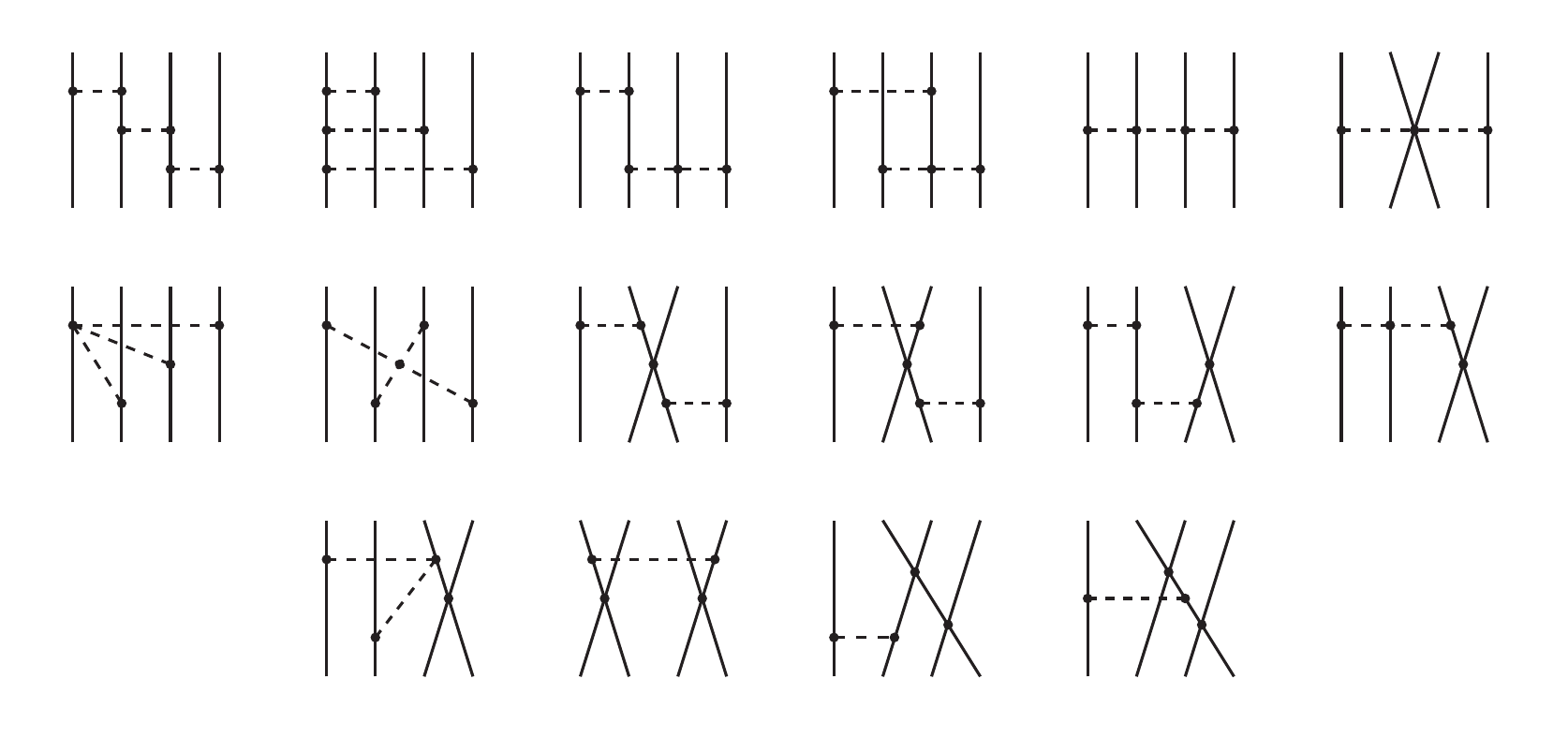}}
\vspace*{-0.5cm}
\caption{Leading four-nucleon force at N$^3$LO.}
\label{fig_4nf_n3lo}
\end{figure}

For connected ($C=1$) $A=4$ diagrams, Eq.~(\ref{eq_nu}) yields
\begin{equation}
\nu = 4 + 2L + 
\sum_i \Delta_i \,.
\label{eq_nu4nf}
\end{equation}
%HERE8
We then see that the first (connected) non-vanishing 4NF is generated at $\nu = 4$ (N$^3$LO), with                   
all vertices of leading type, Fig.~\ref{fig_4nf_n3lo}. 
This 4NF has no loops and introduces no novel parameters~\cite{Epe07}.

For a reasonably convergent series, terms                      
of order $(Q/\Lambda_\chi)^4$ should be small and, therefore,              
chiral 4NF contributions are expected to be very weak.    
This has been confirmed in calculations of the energy of              
$^4$He~\cite{Roz06}
and neutron matter~\cite{Kru13}.

The effects of the leading chiral 4NF in symmetric nuclear matter and pure neutron matter have been
worked out by Kaiser {\it et al.}~\cite{Kai12,KM16}.

\section{Phase IV (2020 -- 2050?) Future: EFT Based Models(?)}
\label{sec_IV}

One of the most fundamental aims in theoretical nuclear physics is to understand 
nuclear structure and reactions in terms of the basic forces between nucleons.
Research pursuing this goal has two major ingredients: nuclear forces and quantum
many-body theory.

As demonstrated in the previous section, the advantage of chiral EFT is that it generates the forces needed (two- and many-body forces) on an equal footing and in a systematic way (namely, order by order), while at the same time improving the  precision of the predictions
and allowing for reliable uncertainty quantifications (cf.\ Sec.~V of Ref.~\cite{EMN17}).

Concerning the second ingredient, during the past two decades, there has been great progress in the development and refinement of diverse many-body methods. Among them are
the no-core shell model~\cite{BNV13}, coupled cluster theory~\cite{Hag14a}, self-consistent Green's functions~\cite{Som14},
quantum Monte Carlo~\cite{Car15},
  and the
in-medium similarity renormalization group method~\cite{Her16}.

In the pursuit of the fundamental aim, a large number of applications of the chiral $NN$ potentials (usually up to N$^3$LO) together with chiral 3NFs (generally just at NNLO) have been conducted.
 These investigations include few-nucleon 
reactions~\cite{Epe02b,NRQ10,Viv13,Gol14}, structure of light- and medium-mass nuclei~\cite{Hag12a,Hag12b,BNV13,Her13,Hag14a,Sim17,Mor17}, and infinite matter~\cite{HS10,Heb11,Hag14b,Cor13,Cor14,Sam15,MS16}.  Although satisfactory predictions have been obtained in many cases, persistent problems continue to pose serious challenges, such as the well-known `$A_y$ puzzle' of nucleon-deuteron scattering~\cite{EMW02,Epe02b,Viv13,Gol14}.
For intermediate mass nuclei, we are faced with systematic overbinding~\cite{Bin14} and
a ``radius problem''~\cite{Lap16}.
 
 Naturally, one would suspect missing 3NFs as the most likely mechanism to solve the open questions, particularly, since most current calculations include 3NFs only at NNLO.
 However, the 3NFs at N$^3$LO~\cite{Ber08,Ber11}  and N$^4$LO~\cite{KGE12,KGE13} (cf.\ Sec.~\ref{sec_3nfn4lo}) are so terribly complicated that 
 including them all is not a realistic task.
 Thus, it appears that chiral EFT, when pursued consistently, may become unmanageable.
 
 In view of this frustrating situation,
 an alternative culture is emerging
 that breaks with conventional practises. As outlined in Secs.~\ref{sec_pot2} and \ref{sec_3nfs}, in the conventional approach, the $\pi N$ LECs are fixed by the $\pi N$ data,
 the $NN$ contacts by the two-nucleon data, and the 3N LECs by three-nucleon data. All applications of these forces in systems with $A>3$ are then predictions. In the unconventional approach, with which presently some researchers are experimenting (see e.g., Refs.~\cite{Eks15,Car16,Eks17}),
 the 2NFs and 3NFs are treated on the manageable NNLO level, and the $\pi N$, NN, and 3NF LECs are determined simultaneously. Furthermore,
 to ensure better results for intermediate-mass nuclei, the
 binding energies and radii of few-nucleon systems and selected isotopes of oxygen and, in some cases, even carbon are included in the optmization procedure. 
 The inclusion of selected spectra of light nuclei in the optimization is also contemplated.
  These interactions then predict binding energies and
 radii of nuclei up to $^{40}$Ca much better~\cite{Eks15}. However, in most of these models, the $NN$ data are reproduced only at very low energies.
 To justify this approach, it is argued that low-energy nuclear observables
 should be sensitive only to low-energy $NN$ properties. 
 Predictive power and large extrapolations do not
 go together well, as small uncertainties in few-body systems get magnified in heavy nuclei.
 Therefore, in this philosophy~\cite{Eks15},
 light nuclei are ``predicted'' by interpolation, heavy nuclei by modest extrapolation. The spirit of this culture is similar to the one of the nuclear mean-field
 and nuclear density functional theory. However, it needs to be stressed that
 approaches of this kind are {\it models}. 
 
 A dogma of chiral EFT is that  a calculation is only meaningful if it includes {\it all} terms up to the given order (i.e., all 2NF, 3NF, and potentially 4NF contributions up to the given order). 
 However,
 the combination of N$^3$LO 2NFs and NNLO 3NFs has already become common practise.
 In the future, we expect that N$^4$LO 2NFs are combined with selected (but not all) N$^4$LO 3NF (contact) terms.
 Skillfully chosen
  combinations carry the potential
 to solve some puzzles that have been oustanding for 30 years, like the $Nd$ $A_y$  puzzle~\cite{Gir16},
 and problems with some spectra of light and intermediate-mass nuclei.
 Since the 3NF terms at N$^3$LO and N$^4$LO are so numerous, selecting just a few of them that show  potential will be the best one can do for a while (if not for ever).
 
 Thus, chiral EFT may ultimately suffer the same fate as meson theory. As explained in 
 Sections~\ref{sec_pions} and \ref{sec_mesons}, meson theory was originally (Phase I) 
 designed to be a quantum field {\it theory}, but later (Phase II) had to be demoted to the level of a {\it model} (a very successful model, though).
 During the current Phase III, the main selling point has been that chiral EFT is a {\it theory} and not just a model and, therefore, its dogmatic use has been pushed. However,
 in analogy to what historically happened to meson theory, during the next phase (namely, Phase IV), we may have to resign ourselves to {\it chiral EFT based models} (that may potentially have great success).
 
 The history of nuclear forces clearly shows a pattern of 30-year phases. Whether these cycles will
 go on forever or whether Phase IV will be the last one, we will know Anno Domini 2050.

\section*{Acknowledgements}
This research is supported in part by the US Department of Energy under Grant No.\ DE-FG02-03ER41270.

\end{document}